\documentclass[10pt,conference]{IEEEtran}

\newcommand{\thetitle}{DockerMock: Pre-Build Detection of Dockerfile Faults through Mocking Instruction Execution}

\usepackage{makecell}
\usepackage{pifont}
\newcommand{\cmark}{\ding{51}}
\newcommand{\xmark}{\ding{55}}

\usepackage{algorithm}
\usepackage[
]{algpseudocode}

\usepackage{amsmath}
\usepackage{wasysym}
\usepackage{amsthm}
\usepackage{mathtools}
\usepackage{balance}
\usepackage{booktabs} 
\usepackage[hidelinks]{hyperref}
\usepackage{cite}
\usepackage{cleveref}
\Crefname{figure}{Fig.}{Fig.}
\usepackage{graphicx}
\usepackage[utf8]{inputenc}
\usepackage{listings}
\usepackage{multirow}
\ifCLASSOPTIONcompsoc
  \usepackage[caption=false,font=normalsize,labelfont=sf,textfont=sf]{subfig}
\else
  \usepackage[caption=false,font=footnotesize]{subfig}
\fi

\usepackage{tikz}
\usetikzlibrary{arrows,calc,patterns,shapes}
\usepackage{pgfplots}
\pgfplotsset{compat=1.16}
\usetikzlibrary{
	pgfplots.dateplot,
	positioning
}
\makeatletter
\pgfdeclareshape{datastore}{
	\inheritsavedanchors[from=rectangle]
	\inheritanchorborder[from=rectangle]
	\inheritanchor[from=rectangle]{center}
	\inheritanchor[from=rectangle]{base}
	\inheritanchor[from=rectangle]{north}
	\inheritanchor[from=rectangle]{north east}
	\inheritanchor[from=rectangle]{east}
	\inheritanchor[from=rectangle]{south east}
	\inheritanchor[from=rectangle]{south}
	\inheritanchor[from=rectangle]{south west}
	\inheritanchor[from=rectangle]{west}
	\inheritanchor[from=rectangle]{north west}
	\backgroundpath{
		\southwest \pgf@xa=\pgf@x \pgf@ya=\pgf@y
		\northeast \pgf@xb=\pgf@x \pgf@yb=\pgf@y
		\pgfpathmoveto{\pgfpoint{\pgf@xa}{\pgf@ya}}
		\pgfpathlineto{\pgfpoint{\pgf@xb}{\pgf@ya}}
		\pgfpathmoveto{\pgfpoint{\pgf@xa}{\pgf@yb}}
		\pgfpathlineto{\pgfpoint{\pgf@xb}{\pgf@yb}}
	}
}
\makeatother
\makeatletter
\newcount\dirtree@lvl
\newcount\dirtree@plvl
\newcount\dirtree@clvl
\def\dirtree@growth{%
	\ifnum\tikznumberofcurrentchild=1\relax
	\global\advance\dirtree@plvl by 1
	\expandafter\xdef\csname dirtree@p@\the\dirtree@plvl\endcsname{\the\dirtree@lvl}
	\fi
	\global\advance\dirtree@lvl by 1\relax
	\dirtree@clvl=\dirtree@lvl
	\advance\dirtree@clvl by -\csname dirtree@p@\the\dirtree@plvl\endcsname
	\pgf@xa=0.5cm\relax
	\pgf@ya=-0.5cm\relax
	\pgf@ya=\dirtree@clvl\pgf@ya
	\pgftransformshift{\pgfqpoint{\the\pgf@xa}{\the\pgf@ya}}%
	\ifnum\tikznumberofcurrentchild=\tikznumberofchildren
	\global\advance\dirtree@plvl by -1
	\fi
}

\tikzset{
	dirtree/.style={
		growth function=\dirtree@growth,
		every node/.style={anchor=north},
		every child node/.style={anchor=west},
		edge from parent path={(\tikzparentnode\tikzparentanchor) |- (\tikzchildnode\tikzchildanchor)}
	}
}
\makeatother

\title{\thetitle}
\author{
	\IEEEauthorblockN{Mingjie Li\IEEEauthorrefmark{1}, Xiaoying Bai\IEEEauthorrefmark{2},
		Minghua Ma\IEEEauthorrefmark{1}, Dan Pei\IEEEauthorrefmark{1}}
	\IEEEauthorblockA{\IEEEauthorrefmark{1}Tsinghua University, China\\
		\{lmj18@mails., mmh16@mails., peidan@\}tsinghua.edu.cn}
	\IEEEauthorblockA{\IEEEauthorrefmark{2}Advanced Institute of Big Data, China\\
		bai\_xiaoying@sina.cn}
}

\theoremstyle{definition}

\newtheorem*{assumption}{Assumption}

\newcommand{\GitHubSampleSize}{488}
\newcommand{\RecentGitHubSampleSize}{294}
\newcommand{\RecentFailedGitHubSampleSize}{120}
\newcommand{\RecentGitHubSampleFailureRate}{40.8\%}
\newcommand{\RecentGitHubDockerfileSize}{80}

\newcommand{\RecentFailedGitHubDockerfileSize}{79}
\newcommand{\GitHubFailureSize}{116}
\newcommand{\GitHubSubmoduleFailureSize}{6}
\newcommand{\GitHubARGFailureSize}{4}
\newcommand{\GitHubFaultSize}{106}

\newcommand{\GitHubDatasetFaultSize}{53}
\newcommand{\GitHubDatasetDockerfileSize}{41}
\newcommand{\GitHubDockerfileFaultRate}{50.0\%}
\newcommand{\GitHubOuterFileNotFoundRate}{60.4\%}
\newcommand{\DockerMockGitHubFalsePositive}{3}
\newcommand{\DockerMockGitHubPrecision}{92.5}
\newcommand{\DockerMockGitHubRecall}{69.8}
\newcommand{\hadolintGitHubPrecision}{1.0}
\newcommand{\hadolintGitHubRecall}{3.8}

\newcommand{\BuildKitGitHubPrecision}{100.0}
\newcommand{\BuildKitGitHubRecall}{64.2}

\newcommand{\hadolintGitHubExecution}{582}
\newcommand{\hadolintGitHubAvgExecutionTime}{0.33 s}
\newcommand{\DockerMockGitHubExecution}{450}
\newcommand{\DockerMockGitHubAvgExecutionTime}{0.18 s}
\newcommand{\BuildKitGitHubBuildTimes}{74}
\newcommand{\hadolintGitHubBuildTimes}{73}

\newcommand{\DockerMockGitHubBuildTimes}{47}
\newcommand{\DockerMockGitHubSavedBuildTimes}{36.5}

\newcommand{\GitHubDatasetTimedRepoSize}{34}
\newcommand{\BuildKitGitHubBuildTime}{2,485.5 s}

\newcommand{\BuildKitGitHubBuildTimeNocache}{3,454.5 s}

\newcommand{\hadolintGitHubSavedTimeRate}{5.2}

\newcommand{\hadolintGitHubSavedTimeRateNocache}{9.4}

\newcommand{\DockerMockGitHubSavedTime}{215.3 s}
\newcommand{\DockerMockGitHubSavedTimeRate}{8.7}

\newcommand{\DockerMockGitHubSavedTimeNocache}{348.9 s}
\newcommand{\DockerMockGitHubSavedTimeRateNocache}{10.1}

\newcommand{\StudentSampleFaultSize}{405}
\newcommand{\StudentSampleRepoSize}{39}
\newcommand{\StudentSampleDockerfileSize}{337}

\newcommand{\StudentDatasetFaultSize}{130}
\newcommand{\StudentDatasetRepoSize}{26}
\newcommand{\StudentDatasetDockerfileSize}{105}
\newcommand{\StudentDockerfileFaultRate}{32.1\%}
\newcommand{\StudentOuterFileNotFoundRate}{45.4\%}
\newcommand{\StudentDatasetTimedSize}{102}
\newcommand{\StudentDatasetBuildTime}{4,484.5 s}
\newcommand{\DockerMockStudentSavedTime}{2,853.2 s}
\newcommand{\DockerMockStudentSavedTimeRate}{63.6}

\newcommand{\DockerMockStudentPrecision}{98.9}
\newcommand{\DockerMockStudentRecall}{66.2}

\newcommand{\hadolintStudentSavedTimeRate}{6.3}

\newcommand{\hadolintStudentPrecision}{6.8}
\newcommand{\hadolintStudentRecall}{9.2}

\newcommand{\BuildKitStudentSavedTimeRate}{50.1}

\newcommand{\BuildKitStudentPrecision}{100.0}
\newcommand{\BuildKitStudentRecall}{56.9}

\begin{document}


\maketitle
\setcounter{page}{1}
\thispagestyle{plain}
\pagestyle{plain}

\begin{abstract}

	Continuous Integration (CI) and Continuous Deployment (CD) are widely adopted in software engineering practice.
	In reality, the CI/CD pipeline execution is not yet reliably continuous because it is often interrupted by Docker build failures.
	However, the existing \textit{trial-and-error} practice to detect faults is time-consuming.
	To timely detect Dockerfile faults, we propose a context-based pre-build analysis approach, named \textit{DockerMock}, through mocking the execution of common Dockerfile instructions. A Dockerfile fault is declared when an instruction conflicts with the approximated and accumulated running context.
	By explicitly keeping track of whether the context is fuzzy, \textit{DockerMock} strikes a good balance of detection precision and recall.
	We evaluated DockerMock with \GitHubDatasetFaultSize{} faults in \GitHubDatasetDockerfileSize{} Dockerfiles from open source projects on GitHub and \StudentDatasetFaultSize{} faults in \StudentDatasetDockerfileSize{} Dockerfiles from student course projects.
	On average, DockerMock detected 68.0\% Dockerfile faults in these two datasets.
	While baseline hadolint detected 6.5\%, and baseline BuildKit detected 60.5\% without instruction execution.
    In the GitHub dataset, DockerMock reduces the number of builds to \DockerMockGitHubBuildTimes{}, outperforming that of hadolint (\hadolintGitHubBuildTimes{}) and BuildKit (\BuildKitGitHubBuildTimes{}).
\end{abstract}

\section{Introduction}

Nowadays, container-based Continuous Integration (CI)~\cite{FowlerCI} and Continuous Deployment (CD)~\cite{FowlerCD} is a prevalent practice to speed up software development and deployment~\cite{Parnin:2017}.
Docker is the most popular container solution~\cite{Flexera:2020}.
However, the CI/CD pipeline execution is often blocked by Docker build failures due to the discrepancy between the local environment and that inside a Docker image.
Previous work found that 17.8\% of historical Docker builds failed in the sample open-source projects~\cite{Wu:2020}.
In the community of deep learning, it was reported that 39.4\% of the jobs based on Docker failed because of accessing a non-existent file or directory~\cite{Zhang:2020}.
In our sample of GitHub projects, \RecentFailedGitHubSampleSize{} out of \RecentGitHubSampleSize{} projects failed to be built, with a failure rate of \RecentGitHubSampleFailureRate{}.
Thus, it is of vital importance to tackle Docker build failures.

\begin{figure}[tb]
	\centering
	\subfloat[The context before line \ref{lst:wrong-path:copy}]{\label{fig:scenario:file:context}
		\begin{minipage}[b]{0.9\columnwidth}
			\resizebox{\linewidth}{!}{
				\begin{tikzpicture}[dirtree]
	\node (variable) {Variables};
	\node [xshift=0.5cm, yshift=-0.4cm] at (variable) {\begin{tabular}{|l|l|}
			\hline
			USER & root \\
			\hline
			PWD & / \\
			\hline
		\end{tabular}
	};
	\node (root) [xshift=1.2cm, yshift=-1.1cm] at (variable) {};
	\node (env)[xshift=1cm, yshift=-2cm]  at (variable) {Environment Variables};
	\node [xshift=0.5cm, yshift=-0.4cm]
		at (env) {\begin{tabular}{|l|l|}
			\hline
			NODE\_VERSION & 12 \\
			\hline
			HOME & /root \\
			\hline
		\end{tabular}
	};
	\node (container) [text width=2cm, xshift=5cm, yshift=0.25cm]
		at (variable) {Files in the \\ Container};
	\node (target) [xshift=-0.9cm, yshift=-0.4cm] at (container) {/}
		child { node {bin}}
		child { node {boot}}
		child { node {etc}}
		child { node {$\cdots$}};
	\node (workspace) [text width=3cm, xshift=2.5cm, yshift=0.45cm] at (container) {Files under the \\ Workspace};
	\node [xshift=-0.9cm, yshift=-0.45cm] at (workspace) {.}
		child { node {$\cdots$}}
		child { node {Dockerfile}}
		child { node {myapp}
			child { node {$\cdots$} }
			child { node {\color{red}{package.json}} }
			child { node {\color{blue}{package-lock.json}} }
		};

	\draw [->,thick] (root.east) to [in = 180, out = 0] (target.west);
\end{tikzpicture}
			}
		\end{minipage}
	}

	\subfloat[Dockerfile example. Both of ``package.json'' at line \ref{lst:wrong-path:copy} and ``package-lock.json'' at line \ref{lst:wrong-path:copy:lock} lack a prefix of ``myapp/''.]{\label{fig:scenario:file:content}
		\begin{minipage}[b]{0.8\columnwidth}
			\scriptsize 
\begin{lstlisting}[
			basicstyle={\ttfamily\color{black}},
			breaklines=true,
			language=bash,
			showspaces=false,
			escapechar=!,
			numbers=left,
			]
FROM node:12

!\color{red}{COPY package.json /app/package.json}\tikz[remember picture] \node [] (point) {};\label{lst:wrong-path:copy}!
!\color{blue}{COPY package-lock.json /app/package-lock.json}\label{lst:wrong-path:copy:lock}!
WORKDIR /app
RUN npm install!\label{lst:wrong-path:npm}!
COPY myapp /app

ENV PORT 80
EXPOSE 80
CMD npm start
\end{lstlisting}
		\end{minipage}
	}
	\caption{Dockerfile example and the context before line \ref{lst:wrong-path:copy} is executed.
	}\label{fig:scenario:file}
\end{figure}

\texttt{docker build}~\cite{dockerbuild} is a command that interprets a Docker \textit{program} called  \textbf{Dockerfile} (see \Cref{fig:scenario:file:content} for an example) written in a Unix-shell-like script language.
A Dockerfile  consists of \textbf{instructions} (written in capitals, \textit{e.g.}, FROM, COPY, RUN) and corresponding parameters and may also embed shell scripts.
A \textbf{fault} (or bug) in a Dockerfile will cause a \textbf{Docker build failure} (or \textbf{failure} in short hereafter).
For example, among \GitHubFaultSize{} failures in our sample, \GitHubDockerfileFaultRate{} are due to Dockerfile faults.
\Cref{fig:scenario:file} shows a Dockerfile example that fails to build. 
Comparing the path specified at line \ref{lst:wrong-path:copy} with the directory structure of source code shows that  ``package.json'' lacking a prefix of ``myapp/'' is a fault. Note that there can be multiple faults in a Dockerfile, \textit{e.g.}, line \ref{lst:wrong-path:copy:lock} in \Cref{fig:scenario:file} has another fault.

\begin{figure*}[tb]
	\centering
	\subfloat[Local trial-and-error]{\label{fig:process:trial-and-error-local}
		\begin{minipage}[b]{0.32\linewidth}
			\includegraphics[width=\linewidth]{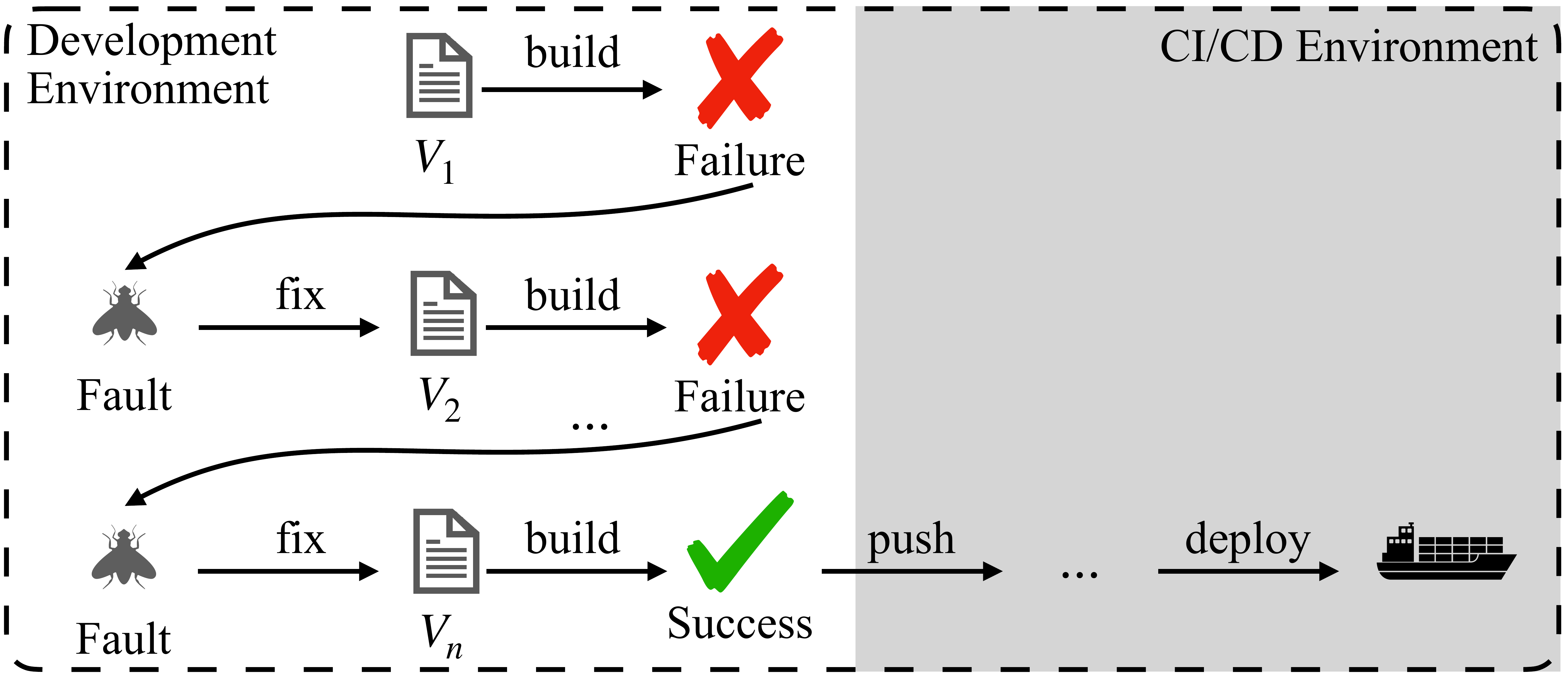}
		\end{minipage}
	}
	\subfloat[Trial-and-error in the CI/CD environment]{\label{fig:process:trial-and-error-ci}
		\begin{minipage}[b]{0.32\linewidth}
			\includegraphics[width=\linewidth]{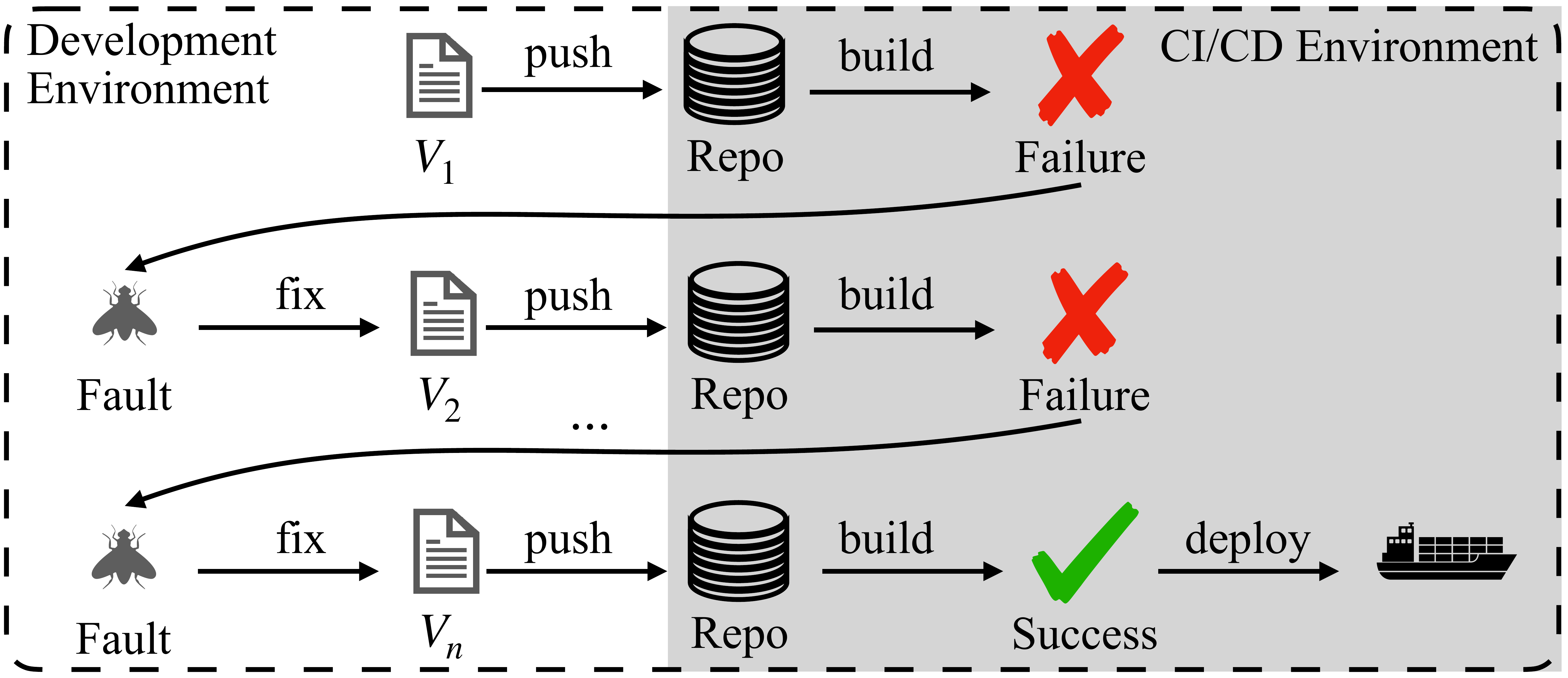}
		\end{minipage}
	}
	\subfloat[Trial-and-error aided by pre-build checking]{\label{fig:process:pre-build}
		\begin{minipage}[b]{0.32\linewidth}
			\includegraphics[width=\linewidth]{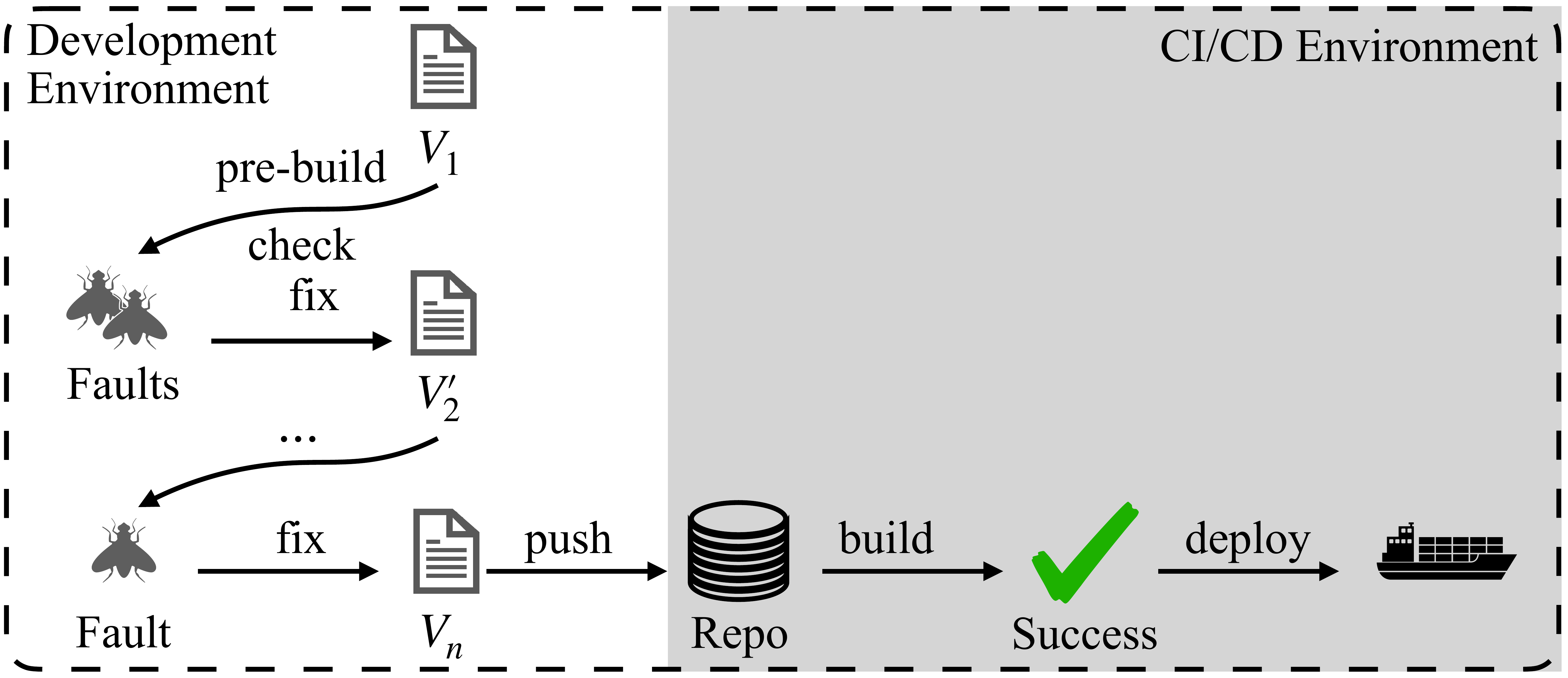}
		\end{minipage}
	}
	\caption{Trial-and-error approach}
\end{figure*}

We focus on studying Dockerfile fault detection.
Previously, developers typically fix the corresponding fault once a build failure happens.
They build again until all faults of the Dockerfile are fixed, named \textit{a trial-and-error approach}.
Such an approach may be applied in the local development environment (\Cref{fig:process:trial-and-error-local}) or the CI/CD cloud environment (\Cref{fig:process:trial-and-error-ci}). 
With the development of cloud computing, more and more build actions are transferred to the cloud~\cite{Zhang:2020}. 
According to our survey in a Software Engineering (SE) course in Fall 2019, 70\% of 96 students relied on the CI/CD pipeline to debug their Dockerfiles.
Using the trial-and-error approach locally can waste time to build.
However, it may suffer more in the CI/CD environment.

More specifically, the trial-and-error approach can waste time because of the large number of builds.
In a sample dataset in \cite{Cito:2017}, it can take ``up to 1539 seconds and on average 90.5 seconds'' to experience the failure that exposes the \textit{first} fault in a Dockerfile.
In practice, each fault may take several trials and builds to be fixed.
Compared to \Cref{fig:process:trial-and-error-local}, each failed Docker build in \Cref{fig:process:trial-and-error-ci} takes extra time for pushing the update to the repository.
Besides, a shared CI server in the cloud may have several tasks, which inevitably suffer more queuing time.

To efficiently detect Dockerfile faults, we introduce a \textbf{pre-build} analysis to reduce the number of builds, inspired by static code analysis for a general-purpose language.
Before the Dockerfile is built locally or pushed to the CI server, pre-build checking aims to detect as many faults as possible to avoid build failures, as shown in \Cref{fig:process:pre-build}.
Thus, the developers can directly fix the detected faults in the development environment, instead of waiting for minutes to tens of minutes of Docker build in the bare trial-and-error approach.

Despite the potential of pre-build analysis, there are few such tools available so far.
Dockerfile Linter (hadolint)~\cite{dockerfile:linter} parses a Dockerfile and inspects the syntax tree against best practices without considering any other part of the project.
Therefore, hadolint can only detect ``code smells" in a Dockerfile program~\cite{Schwarz:2018}, but just a small fraction of code smells are faults.
RUDSEA~\cite{Hassan:2018} utilizes the change of environment-related code scope to recommend Dockerfile updates, preventing more Dockerfile faults.
However, the faults that can be avoided by RUDSEA are also a subset of real-world ones.
Enabled by BuildKit~\cite{buildkit}, Docker can prepare instructions concurrently.
Therefore, a fault, such as a nonexistent source file specified by COPY, may be discovered earlier, even though the instruction has not been executed.
However, the common shortcoming of these tools is that they can only detect limited types of faults.
The undetected failures continue wasting time, which can potentially be detected in a pre-build manner.

In this paper, we aim to develop a pre-build analysis tool to efficiently cover as many types of Dockerfile faults as possible.
To this end, we first propose a taxonomy for Docker build faults and then conduct an empirical study on Dockerfile faults based on two datasets that we collected and labeled.
One dataset contains \GitHubDatasetDockerfileSize{} Dockerfile from GitHub samples, and another contains \StudentDatasetDockerfileSize{} Dockerfiles from \StudentDatasetRepoSize{} student course project repositories.
We employed the trial-and-error approach to obtain the ground truth of the fault type labels.
In other words, we manually run \texttt{docker build} to find, fix, and label each fault in each of the Dockerfile in these two datasets.
We publish these two labeled Dockerfile fault datasets to serve as benchmark datasets for future research, the first ones of their kind, to the best of our knowledge.
One key observation in our empirical study based on the above two datasets is that \textit{it is necessary to keep track of the \textbf{context} (see \Cref{fig:scenario:file:context} for an example) of the Dockerfile program as \texttt{docker build} interprets each line of instructions}, for pre-build analysis to cover many Dockerfile faults.
In retrospect, this observation is intuitive in that the Dockerfile program interacts with the operating system through environment variables, files, and shell commands. Thus, its running context does affect whether a potential fault takes effect or not.

Therefore, \textit{the core idea of our proposed approach, called \textbf{\textit{DockerMock}}, is a pre-build detection of Dockerfile faults through mocking the \texttt{docker build} execution of the instructions in a Dockerfile}.
A \texttt{docker build} mock keeps track of the context without running \texttt{docker build} or wait minutes to tens of minutes to discover each fault in a Dockerfile.
However, \textbf{the major challenge is how to mitigate the false positives introduced by the imprecise context}.
The information available to a \texttt{docker build} mock, such as our DockerMock, is likely much less than the actual \texttt{docker build}.
Thus, the context maintained by \textit{DockerMock} is only an approximation of the real context maintained by \texttt{docker build}.
As a result, there is a risk that the fault detected by \textit{DockerMock} is not a fault (\textit{i.e.}, we have a  false  positive).

To tackle the above challenge, in addition to keeping track of contexts, \textit{DockerMock} also \textit{explicitly keeps track of whether the context is imprecise (or \textbf{fuzzy}}).
Warnings derived from the above fuzzy context will be dropped, avoiding false positives.
Incorporated with our proposed \textbf{fuzzy processing}, the scope of the fuzzy context can be limited.
We try to cache some information of the base Docker image to bring extra certainty into the analysis.

Our evaluation based on the two benchmark datasets mentioned before shows that, on average, DockerMock detected 68.0\% Dockerfile faults in these two datasets.
While baseline hadolint detected 6.5\% and baseline BuildKit detected 60.5\% without instruction execution.
In the GitHub dataset, the number of builds is reduced to \DockerMockGitHubBuildTimes{} by DockerMock, outperforming that of hadolint (\hadolintGitHubBuildTimes{}) and BuildKit (\BuildKitGitHubBuildTimes{}).
In the student course project dataset,  DockerMock reduces the total build time by \DockerMockStudentSavedTimeRate{}\% compared to the \textit{trial-and-error} approach, outperforming that of hadolint (\hadolintStudentSavedTimeRate{}\%) and BuildKit (\BuildKitStudentSavedTimeRate{}\%).

This paper's major contributions are the following:

\begin{itemize}
    \item This paper is the first systematic study to identify the research problem of \textit{Dockerfile fault detection} in terms of efficiency and effectiveness.
    We label and publish two Dockerfile fault datasets~\cite{dataset}
    that can serve as benchmarks for future Dockerfile research.
    \item We present a taxonomy of Docker build faults and conduct the first empirical study on Dockerfile fault types to the best of our knowledge.
    \item We propose a pre-build analysis approach, named DockerMock~\cite{code}, which incorporates context mocking and fuzzy processing to achieve a good balance between false positives and false negatives. To the best of our knowledge, this is the first attempt to use context and pre-build technique to do Dockerfile fault detection.
\end{itemize}

 The rest of this paper is organized as follows.
 \Cref{sec:pattern} analyzes the fault taxonomy of Dockerfiles.
 \Cref{sec:method} introduces our proposed method DockerMock.
 The implementation detail is further discussed in \Cref{sec:implmentation}.
 \Cref{sec:experiment} evaluates DockerMock with Dockerfiles of the GitHub sample and student projects.
 At the end of this paper, \Cref{sec:related} discusses related works, and \Cref{sec:conclusion} concludes.

\section{Faults Taxonomy and Empirical Study}\label{sec:pattern}

In this section, we describe the background of Docker and Dockerfile.
Then we introduce our two datasets. 
To systematically study Dockerfile faults, we propose a taxonomy on Docker build faults and present an empirical study on Dockerfile faults.

\subsection{Background}

\noindent \textbf{Docker and Dockerfile}: Docker is a prevalent way to release software, which is packed in a Docker image.
A Docker image can be built based on a \textit{Dockerfile}, which consists of \textit{instructions} (written in capitals, \textit{e.g.}, FROM, COPY, RUN) and corresponding parameters.
It may also embed shell scripts in RUN instructions.
While building a Docker image, the base image is instantiated as a \textit{container}, which is similar to a virtual machine.
Following a sequence of instructions recorded in the Dockerfile, environment variables and files in the container are changed, as well as other attributes configuring the software, such as the working directory.

\noindent \textbf{Context}:
A collection of changeable attributes is named as the \textit{context} for an instruction, such as environment variables and files.
Note that the meaning of the context in Docker documentation~\cite{dockerbuild} is a subset in our scenario.
The context is delivering and accumulating the effect of previous instructions.
For example, at the end of the Dockerfile example \Cref{fig:scenario:file}, context is expected to have an environment variable \texttt{PORT} of ``80'', the working directory of ``/app'', and files used to run this project under ``/app'', based on Docker image \texttt{node:12}.

\noindent \textbf{Failure}: Failure is one kind of Docker build result when the Docker image fails to be built.

\noindent \textbf{Fault}: A fault is the cause of a failure, discovered within the project.
In this paper, we focus on Dockerfile faults. 


\subsection{Data Set}\label{sec:experiment:data}


%

Based on a Dockerfile, the \texttt{docker build} command may not build a Docker image successfully, which is referred to as a \textit{failure}.
For example, 560 GitHub projects in the dataset~\cite{Cito:2017} include 203 Dockerfiles that failed to be built.
More specifically, 144 of them are out of updates since January 2017, and 42 failures are caused by unavailable base images.
These failures caused by lack of maintenance are not the focus in this paper, as they result from external reasons.
Many failures, however, result from intrinsic \textit{faults} that are internal reasons.

We first collect Docker projects from GitHub (a popular open-source community) and a SE course that contains available details about the developing process.
Then, Dockerfiles are built on a Linux server with an x86-64 architecture.
We further do our best to fix build failures to discover all the faults in each Dockerfiles.
The insights gained on the two datasets can be mutually confirmed.

\begin{table}[tb]
	\centering
	\caption{Summary of \texttt{docker build} faults}\label{tab:classification}
	\scriptsize
\begin{tabular}{llrr}
    \toprule
    Position & Pattern & \# $\mathcal{D}_{G}$ & \# $\mathcal{D}_{S}$ \\
    \midrule
    \multicolumn{2}{l}{Source} & 7 & 202 \\
    \multicolumn{2}{l}{Package Manage} & 46 & 73 \\
    \multicolumn{2}{l}{Dockerfile} & \textbf{53} & \textbf{130} \\
    & Syntax-Mistake & 0 & 3 \\
    & Instruction-Misuse & 0 & 8 \\
    & Command-Misuse & 7 & 5 \\
    & Command-Not-Found & 4 & 9 \\
    & Outer-File-Not-Found & 32 & 59 \\
    & Inner-File-Not-Found & 3 & 37 \\
    & Image-Not-Found & 2 & 9 \\
    & Image-Version-Mismatch & 3 & 0 \\
    & Permission-Denied & 1 & 0 \\
    & Require-Manual-Input & 1 & 0 \\
    \midrule
    \multicolumn{2}{l}{All} & 106 & 405 \\
    \bottomrule
\end{tabular}

\end{table}

\subsubsection[GitHub Sample]{GitHub Sample \rm{($\mathcal{D}_{G}$)}}

We follow the steps of \cite{Cito:2017} to obtain a sample of GitHub projects.
In February 2020, we retrieved 73,709 repositories that contain a file with a path of ``Dockerfile'' from Google BigQuery\footnote{https://bigquery.cloud.google.com/github}.
We queried GitHub API and filtered 28,975 projects which have been updated since 2018.
Limited by the manual effort of fixing and labeling failures, 500 repositories were sampled, 12 of which were removed from the sample because the Dockerfile no longer exists or the hash of it is duplicated.
We built the default branch of the remained \GitHubSampleSize{} projects.
Among \RecentGitHubSampleSize{} repositories that have at least one commit since 2018, we failed to build Dockerfiles of \RecentFailedGitHubSampleSize{} projects, with a failure rate of \RecentGitHubSampleFailureRate{}.
Dockerfiles of \RecentGitHubDockerfileSize{} projects have been updated since 2018.
\GitHubFailureSize{} failures are collected in \RecentFailedGitHubDockerfileSize{} Dockerfiles, except one Dockerfile, which fails in build intentionally\footnote{https://github.com/voxpupuli/puppet-windowsfeature}.
\GitHubSubmoduleFailureSize{} failures are fixed via initializing git submodules.
\GitHubARGFailureSize{} failures arise due to the lack of default value for the ARG instructions, which have to be specified while building the Docker image.
The rest \GitHubFaultSize{} faults are summarized in the `` \# $\mathcal{D}_{G}$'' column of \Cref{tab:classification},
including \GitHubDatasetFaultSize{} Dockerfile faults from \GitHubDatasetDockerfileSize{} GitHub projects.

\subsubsection[Student Projects]{Student Projects \rm{($\mathcal{D}_{S}$)}}
In a SE course in 2019, 228 students formed into 54 teams, developing using Docker in a DevOps manner.
The project developing process lasted 2 months.
The course provided example repositories with Dockerfile and CI configuration as reference for students.
Except teams using VHDL or Verilog, 1,150 CI jobs in 86 repositories from 48 teams failed in the stage ``build''.
We sampled 22 teams and collected \StudentSampleFaultSize{} faults in \StudentSampleDockerfileSize{} versions from \StudentSampleRepoSize{} repositories.
The ``\# $\mathcal{D}_{S}$'' column of \Cref{tab:classification} shows the summary of those faults.
As a data set derived from student projects,  $\mathcal{D}_{S}$ consists of \StudentDatasetFaultSize{} Dockerfile faults, 
related to \StudentDatasetRepoSize{} repositories and \StudentDatasetDockerfileSize{} git commits.

\begin{table*}[tb]
	\caption{Dockerfile fault types. For a given method, a type can be fully(\cmark) / partially(P) detected or cannot(\xmark) be detected. The statement of a certain ability may be not available(N/A).}\label{tab:patterns}
	\begin{tabular}{p{0.16\linewidth}p{0.45\linewidth}cccc}
		\toprule
		Type & Description & hadolint & BuildKit & RUDSEA & DockerMock \\
		\midrule
		Syntax-Mistake & The Dockerfile or the embedded shell script fails to be parsed. & \cmark & \cmark & N/A & \cmark \\
		Instruction-Misuse & The arguments of a Dockerfile instruction violate its requirements. For example, with multiple source arguments, COPY instruction requires the destination path ends with ``/''. & P & P & N/A & P \\
		Command-Misuse & The arguments of a shell command violate its requirements. For example, \texttt{cp} requires the \texttt{-r} option while copying a directory. & \xmark & \xmark & \xmark & P \\
		Command-Not-Found & An unavailable shell command is invoked, for example, executing \texttt{java} in the latest python image. & P & \xmark & \xmark & P \\
		Outer-File-Not-Found & A referred file does not exist in the workspace. & \xmark & \cmark & \cmark & \cmark \\
		Inner-File-Not-Found & A referred file does not exist in the container's context. & \xmark & \xmark & P & P \\
		Image-Not-Found & The base Docker image is not accessible. & \xmark & \cmark & \xmark & \xmark \\
		Image-Version-Mismatch & A package manager may pin the version of the compiler, which can be incompatible with the tagged image version. & \xmark & \xmark & \cmark & \cmark \\
		Permission-Denied & Some commands require the privilege of the root user, while some other commands cannot be executed as root. & \xmark & \xmark & \xmark & \xmark \\
		Require-Manual-Input & A command may require confirmation, which blocks the build process. & P & \xmark & \xmark & P \\
		\bottomrule
	\end{tabular}
\end{table*}

\subsection{Docker Build Faults Taxonomy}


We collect two datasets (\cref{sec:experiment:data}) and apply open coding~\cite{Pandit:1996} to classify faults manually.
Faults can be divided into three groups according to where the fault is located.
The fault type distribution in two datasets is shown in \Cref{tab:classification}.

\paragraph{Source Code Faults}

If the application is compiled and packed while building a Docker image, a fault hiding in the source code will lead to a failure.

\paragraph{Dependency Faults}

Another kind of common fault is related to package management.
There can be faults in the configuration file for a package manager, such as a syntax mistake or an incomplete dependency list.

\paragraph{Dockerfile Faults}

The Dockerfile has to be changed to fix such a fault, which take up \GitHubDockerfileFaultRate{}, \StudentDockerfileFaultRate{} in $\mathcal{D}_{G}$ and $\mathcal{D}_{S}$, respectively.

\begin{figure*}[tb]
    \centering
	\begin{tikzpicture}
	\small
	\begin{axis}[
	ybar,
	symbolic x coords={RUN,COPY,ENV,FROM,WORKDIR,EXPOSE,CMD,COMMENT,Others},
	xtick=data,
	xlabel=Instruction,
	ylabel=Percentage (\%),
	nodes near coords,
	nodes near coords align={vertical},
	width=\linewidth,
	height=0.3\linewidth,
	]
	\addplot [pattern=north east lines] table[x=instruction,y=dg,col sep=comma]{dat/instruction-percentage.csv};
	\addplot [pattern=none] table[x=instruction,y=ds,col sep=comma]{dat/instruction-percentage.csv};
	\addplot [pattern=horizontal lines] table[x=instruction,y=github,col sep=comma]{dat/instruction-percentage.csv};
	\legend{$\mathcal{D}_{G}$,$\mathcal{D}_{S}$,GitHub~\cite{Cito:2017}}
	\end{axis}
	\end{tikzpicture}
	\caption{Percentages of Dockerfile instructions.
		The percentage of COPY instruction for GitHub projects counts ADD as well.
		``Others'' includes ARG, ENTRYPOINT, HEALTHCHECK, LABEL, MAINTAINER, USER, and VOLUME instructions.}\label{fig:compare:instruction}
\end{figure*}
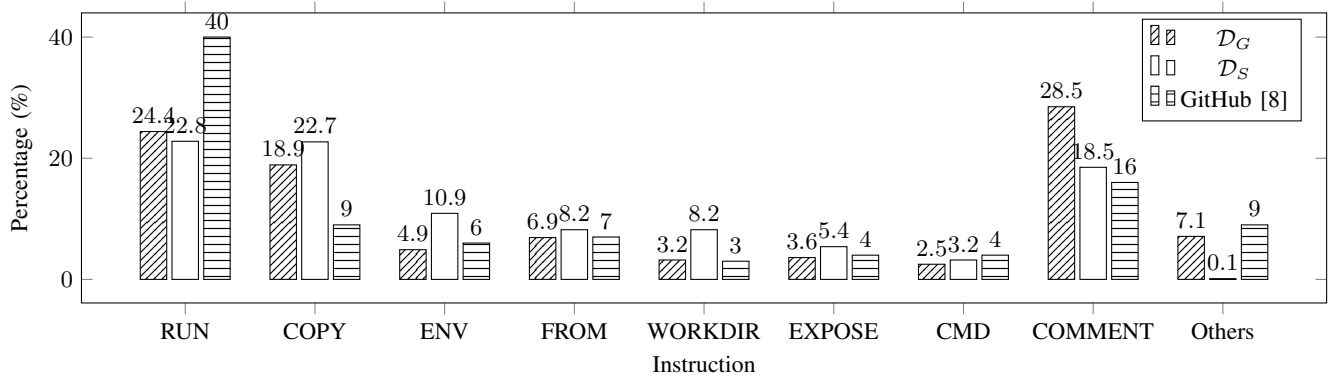

\subsection{Empirical Study on Dockerfile Faults}

The faults related to both source code and package management are expected to be detected via traditional testing, and this paper focuses on Dockerfile faults.

\begin{description}
	\item[\textbf{RQ1}] What are the patterns of faults in Dockerfiles?
\end{description}

\subsubsection{Dockerfile fault types}

We carefully summarize Dockerfile faults into ten types in \Cref{tab:patterns} and the fault type distribution in two datasets is shown in \Cref{tab:classification}.
Outer-File-Not-Found is the most frequent fault type in both of the two datasets.
If a file referred by the source parameter of COPY/ADD instruction does not exist in the project, an Outer-File-Not-Found arises.
Image-Version-Mismatch arises if the image specified by FROM does not satisfy the requirement in the configuration file for the package manager.
For example, ``Gemfile'' is the configuration file of \texttt{bundle}, which is a package manager for Ruby.
If ``Gemfile'' specifies ``\texttt{ruby 2.7}'', using the official Docker image \texttt{ruby:2.6} will fail.
While Command-Not-Found is usually triggered by a command explicitly written in the Dockerfile, a missed command can also be invoked implicitly.
For example, ``\texttt{go get}'' needs \texttt{git} to fetch required packages, while \texttt{git} may not be available.
In summary, \textit{it is necessary to keep track of the context of the Dockerfile program as \texttt{docker build} interprets each line of instructions}.

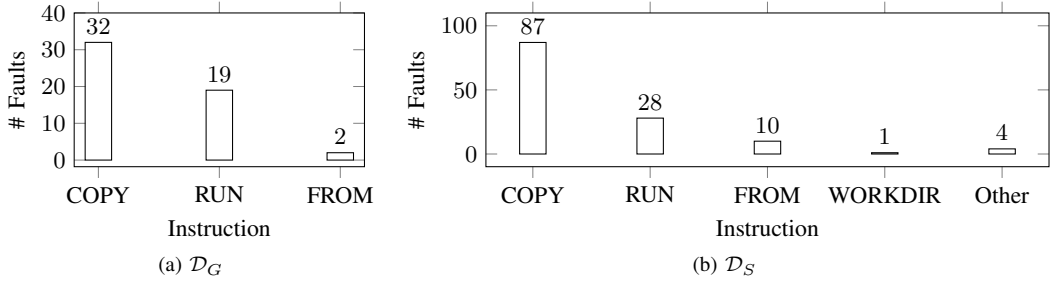
\begin{figure*}[tb]
    \centering
    \subfloat[$\mathcal{D}_{G}$]{\label{fig:data:github:fault-instruction-rate}
	\begin{tikzpicture}
	\small
	\begin{axis}[
	ybar,
	symbolic x coords={COPY,RUN,FROM},
	xtick=data,
	ymax=40,
	xlabel=Instruction,
	ylabel=\# Faults,
	nodes near coords,
	nodes near coords align={vertical},
	width=0.3\linewidth,
	height=0.2\linewidth,
	]
	\addplot [pattern=none] table[x=instruction,y=dg,col sep=comma]{dat/failed-instruction-dg.csv};
	\end{axis}
	\end{tikzpicture}
	}
    \subfloat[$\mathcal{D}_{S}$]{\label{fig:data:student:fault-instruction-rate}
	\begin{tikzpicture}
	\small
	\begin{axis}[
	ybar,
	symbolic x coords={COPY,RUN,FROM,WORKDIR,Other},
	xtick=data,
	ymax=110,
	xlabel=Instruction,
	ylabel=\# Faults,
	nodes near coords,
	nodes near coords align={vertical},
	width=0.5\linewidth,
	height=0.2\linewidth,
	]
	\addplot [pattern=none] table[x=instruction,y=ds,col sep=comma]{dat/failed-instruction-ds.csv};
	\end{axis}
	\end{tikzpicture}
	}
	\caption{\# Faults per kind of instructions. ``Other'' includes faults that arise in no instruction, such as Syntax-Mistake ``unknown instruction''.}
	\label{fig:data:fault-instruction-rate}
\end{figure*}

\begin{figure*}[tb]
    \centering
    \subfloat[$\mathcal{D}_{G}$]{\label{fig:data:github:fault-dockerfile-rate}
        \includegraphics[width=0.4\linewidth]{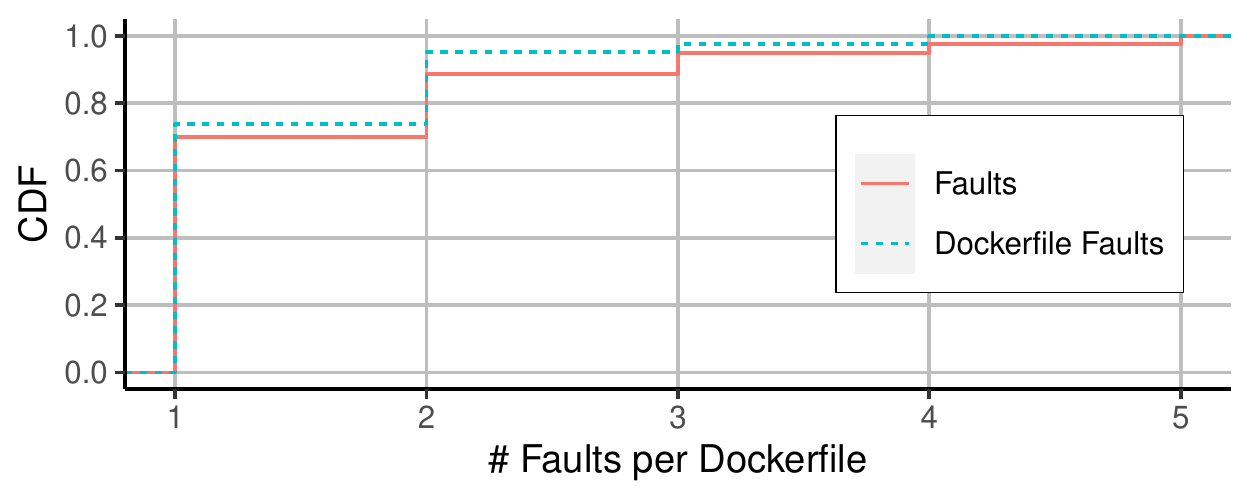}
    }
    \subfloat[$\mathcal{D}_{S}$]{\label{fig:data:student:fault-dockerfile-rate}
        \includegraphics[width=0.4\linewidth]{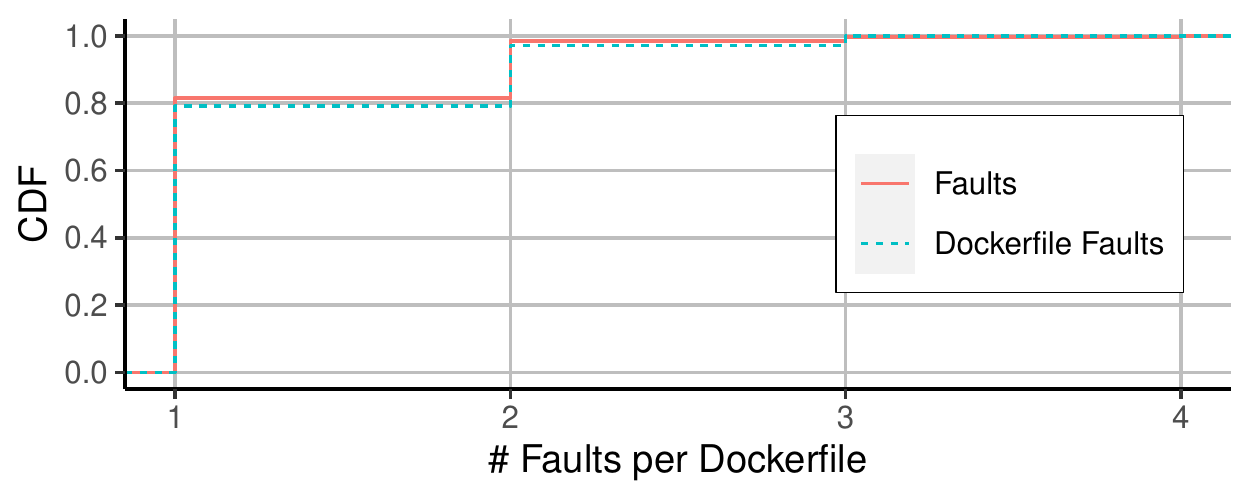}
    }
	\caption{Cumulative distribution function (CDF) of \# faults per Dockerfile}\label{fig:data:fault-dockerfile-rate}
\end{figure*}

\subsubsection{Fault pattern in quantity}

\Cref{fig:compare:instruction} compares the percentages of Dockerfile instructions among $\mathcal{D}_{G}$, $\mathcal{D}_{S}$, and projects in GitHub.
There are more COPY instructions and fewer RUN instructions in Dockerfiles with faults.
\Cref{fig:data:fault-instruction-rate} shows Dockerfile fault distribution among instructions.
COPY and RUN are the most error-prone instructions in both $\mathcal{D}_{G}$ and $\mathcal{D}_{S}$.

\Cref{fig:data:fault-dockerfile-rate} shows how many faults each Dockerfile has in $\mathcal{D}_{G}$ and $\mathcal{D}_{S}$, respectively.
\textit{Around 20\% of Dockerfiles contain more than one Dockerfile faults}.
Via fixing previous faults, another failure may arise.
In \Cref{fig:scenario:file}, the COPY instruction at line \ref{lst:wrong-path:copy:lock} refers to ``package-lock.json'', which is also not under the context.
After the faults in line \ref{lst:wrong-path:copy} and \ref{lst:wrong-path:copy:lock} are fixed, there is no more failure.
Thus, \texttt{npm install} at line \ref{lst:wrong-path:npm} should not be a fault even though ``package.json'' fails to be settled under ``/app'' at original line \ref{lst:wrong-path:copy}.

\section{DockerMock}\label{sec:method}

We present DockerMock, a pre-build fault detector for Dockerfiles, via mocking instruction execution, as shown in \Cref{fig:method:workflow}.
Mocking is a unit testing practice, replacing dependencies with mock objects~\cite{Spadini:2017}.
DockerMock simulates the build process of the Dockerfile and warns the violation between mock context and the requirement of any mock instruction.

\subsection{Workflow}

First, a Dockerfile will be parsed into a syntax tree. The syntax tree of a Dockerfile contains a sequence of instructions. Each instruction has its name, such as RUN, and its parameters. The only parameter of a RUN instruction is a shell script, which will be further parsed according to the shell grammar.

Second, DockerMock will traverse the syntax tree parsed from the Dockerfile, mocking instruction execution.
During the traverse, mock Dockerfile instructions and shell commands interact with mock context (\cref{sec:method:context}).
Warnings will be raised if the mock context does not satisfy the requirement of a mock instruction.
Incomplete context and its analysis is sculptured carefully to balance detection ability and risk of false positives (\cref{sec:method:fuzzy-handling}). 

\begin{figure}[tb]
	\begin{algorithmic}[1]
		\Procedure{DockerMock}{$dockerfile$, $workspace$}
			\State $context \gets \Call{Initialize}{workspace}$
			\For{$instruction \in \Call{Parse}{dockerfile}$}
				\State $result, context \gets \Call{Mock}{instruction, context}$
				\If {$result$ if fuzzy}
					\State Warnings are dropped
					\State \Comment $context$ has been updated during mocking
				\ElsIf{$result$ implies conflict}
					\State Warn the conflict
				\Else \Comment $context$ is updated precisely
				\EndIf
			\EndFor
		\EndProcedure
	\end{algorithmic}
	\caption{Workflow of DockerMock}\label{fig:method:workflow}
\end{figure}

For example,  \Cref{fig:scenario:directory} displays a Dockerfile snippet with a fault.
It will be parsed as three instructions, FROM, COPY, and WORKDIR.
The FROM instruction at the first line will initialize the mock context.
As for the COPY instruction at line \ref{lst:overwrite:copy}, ``db\_initialize.sql'' is found as a regular file in the workspace.
``/database''  is created with the same type of regular file as ``db\_initialize.sql'' because DockerMock fails to find ``/database'' in the mock context.
The WORKDIR instruction at line \ref{lst:overwrite:workdir} requires its parameter ``/database''  to point to a directory, where a failure arises.
Both COPY and WORKDIR instructions are correct if they are taken separately.
However, there is conflict, as they are put in the wrong order.

\begin{figure}[tb]
	\centering
	\begin{minipage}[b]{0.8\linewidth}
		\small
\begin{lstlisting}[
		basicstyle={\ttfamily\color{black}},
		breaklines=true,
		language=bash,
		showspaces=false,
		escapechar=!,
		numbers=left,
		]
FROM node:12
COPY db_initialize.sql /database!\label{lst:overwrite:copy}!
WORKDIR /database!\label{lst:overwrite:workdir}!
\end{lstlisting}
	\end{minipage}
	\caption{Instruction-Misuse of WORKDIR instruction. \texttt{/database} will be a regular file after the COPY instruction at line \ref{lst:overwrite:copy}. Thus, it is illegal to change the working directory as \texttt{/database} at line \ref{lst:overwrite:workdir}.
	}\label{fig:scenario:directory}
\end{figure}

\subsection{Mock Context}\label{sec:method:context}

Context includes different types of attributes.
Variables are fundamental attributes, such as environment variables, while files are more complex attributes than variables.

\textit{Variable} is a general kind of context, managed as a mapping from name to its value.
Some variables are mandatory, and failures arise if missing, such as the base image specified by Dockerfile's FROM instruction.
Optional variables often have their default values.
For example, \textit{environment variables} are such attributes for the shell.
Each environment variable is optional, with a default value of an empty string.

\textit{Files} are often organized as a tree.
A \textit{regular file} can only be a leaf of a file tree, while a \textit{directory} can also have children of files.
There can be several file trees to mock separated files.
For example, \textit{workspace} means the directory contains source code, which shall be organized as a tree.
The workspace is isolated from files in the Docker container described by the Dockerfile, although the previous file tree is usually copied as a sub-tree of the latter one.
The structure of a tree brings more expressive power than a list of file paths, as a directory and its children can be operated at the same time. 
For example, to copy a directory, only the parent of that directory is changed, instead of a bulk of file paths which are the children of the directory.
\textit{Executables} are special regular files, which function as shell instructions.
Given an instruction, the shell will try to treat the name as a path in the file system or locate the executable under the directories listed in the \texttt{PATH} variable.

\subsection{Fuzzy Processing}\label{sec:method:fuzzy-handling}

An attribute is labeled as \textit{fuzzy} if there is only incomplete knowledge of it.
Whether a fuzzy attribute has been set any value is not sure, and its value is not certain.
During the analysis, context becomes fuzzy in some situations.
\begin{itemize}
	\item The context will be initialized to be fuzzy if DockerMock is unaware of the base Docker image specified by a FROM instruction.
	\item If a mock is unavailable, the update of the corresponding instruction to the context is unknown.
	\item DockerMock will not access the network resource during analysis.
	\item After a fault is detected, the Dockerfile shall be fixed, whose effect on the context is unknown.
\end{itemize}

DockerMock treats fuzzy context carefully to strike a balance between false positives and missed faults.
Generally, an execution fails if the exit code is non-zero.
With the fuzzy part of the context accessed, the analysis workflow is confronted with the bias between fuzzy context value and the real one.
The exit code shall be labeled as fuzzy, and warnings shall be dropped, as such a ``failure'' may be a false positive due to imprecise context.
Fuzzy processing is designed to answer the following two questions.
\begin{itemize}
	\item How to update context when an instruction accesses the fuzzy part of the context?
	\item How to update context if execution fails?
\end{itemize}

Mock is the basic unit to handle fuzzy context and execution failure.
If the fuzzy part of the context required by a mock has a limited effect on the overall context, the mock can handle it by itself.
For example, \texttt{npm install} can simply create a fuzzy directory ``node\_modules/'' under the working directory, without fuzzing other parts of the context.
After the execution is detected to be failed, the mock can automatically fix the failure, if the target command is probably unchanged after a fixing.
For example, the mock for \texttt{cp} can continue analysis if just the ``-r'' option is missing.
With a fuzzy execution list, the mock shell interpreter shall still execute the target executable, as introducing a potential missing command is likely a fixing.

Fuzzy processing in mocks depends on the quality of implementation largely, and the strategy may not be as obvious as complementing a missing option.
After a fault is detected, a fixing varies from adding an option to insertion of extra Dockerfile instructions.
Fuzzing the whole context is a more general strategy.
With fuzzy processing, it will also be useful to partially implement a mock, describing which part of context will not be affected.
However, the fuzzy status of a single attribute spreads fast with such a strategy.

To slow the spread of fuzzy status and bring extra certainty into the analysis, we presume that the fuzzy context required by an instruction is also ``correct'', without labeling the whole context as fuzzy.
For example, ``\texttt{mkdir \$HOME/src}'' will try to create a fuzzy directory ``/src'', if the environment variable \texttt{HOME} is fuzzy and treated as an empty string.
Meanwhile, the exit code of the execution shall still be labeled as fuzzy.
If a relative path is adopted and two fuzzy paths share the same fuzzy prefix, this assumption may work well but also bring the risk of false positives at the same time (discussed in \Cref{sec:discussion}).

\begin{assumption}
	The fuzzy context required by an instruction is real.
	However, the exit code of the instruction is still fuzzy.
\end{assumption}

DockerMock can also leverage information in Docker images, called \textit{prior context}.
While mocking a FROM instruction, mock context can be initialized with the real one.
It is expensive to fetch prior context, but it also works for following analysis of the same or other projects.
Meanwhile, it is not operational to collect all the information about a base image.

\section{Implementation}\label{sec:implmentation}

We have implemented DockerMock with Python.
The implementation has been tested in CentOS, macOS, Ubuntu, and Windows.
Morbig~\cite{Regis-Gianas:2018} is a POSIX shell parser, which is used by DockerMock to transform a shell script embedded in a RUN instruction into a concrete syntax tree.

Implemented mocks are listed in \Cref{tab:implement:rule}.
A partially completed mock may label part of the context as fuzzy.
For example, if the first parameter for the mock \textit{npm} is ``install'', only the directory ``node\_modules/'' under the working directory will be labeled as fuzzy, leaving environment variables untouched.
\Cref{fig:result:example} displays an example checking output for the project shown in \Cref{fig:scenario:file}.
Messages labeled as ``WARNING'' or ``ERROR'' point to the fault detected by DockerMock.

\begin{table}[tb]
	\centering
	\caption{Implemented mocks}
	\label{tab:implement:rule}
	\scriptsize
	\begin{tabular}{lll}
		\toprule
		Language & Integrity & Mocks \\
		\midrule
		Shell & Complete & cd, chmod, cp, echo, env, export, mv, mkdir, pwd, rm \\
		& Partial & .(dot), apt, bundle, git, go, ls, npm, pip, python, touch \\ \midrule
		Dockerfile & Complete
		& \makecell[c]{ADD, CMD, COPY, ENTRYPOINT, ENV, FROM, \\ RUN, WORKDIR}\\
		\bottomrule
	\end{tabular}
\end{table}

\begin{figure}[tb]
	\centering
\begin{lstlisting}[
	basicstyle={\scriptsize\ttfamily\color{black}},
	breaklines=true,
	frame=single,
% 	firstline=1,lastline=23, % The last line for .gitlab-ci.yml can be removed
	]
>python -m dockermock --local-dir ./examples/wrong_path
git version 2.24.0
INFO:dockermock.sensor.dockerfile.checker:FROM node:12
INFO:dockermock.sensor.dockerfile.checker:COPY package.json /app/package.json
ERROR:dockermock.sensor.dockerfile.checker:./package.json: No such file or directory
WARNING:dockermock.sensor.dockerfile.checker:The format of COPY shall be

    COPY [--chown=<user>:<group>] <src>... <dest>
    COPY [--chown=<user>:<group>] ["<src>",... "<dest>"]

INFO:dockermock.sensor.dockerfile.checker:COPY package-lock.json /app/package-lock.json
ERROR:dockermock.sensor.dockerfile.checker:./package-lock.json: No such file or directory
WARNING:dockermock.sensor.dockerfile.checker:The format of COPY shall be

    COPY [--chown=<user>:<group>] <src>... <dest>
    COPY [--chown=<user>:<group>] ["<src>",... "<dest>"]

INFO:dockermock.sensor.dockerfile.checker:WORKDIR /app
INFO:dockermock.sensor.dockerfile.checker:RUN npm install
INFO:dockermock.sensor.dockerfile.checker:COPY myapp /app
INFO:dockermock.sensor.dockerfile.checker:ENV PORT 80
INFO:dockermock.sensor.dockerfile.checker:EXPOSE 80
INFO:dockermock.sensor.dockerfile.checker:CMD npm start
\end{lstlisting}
	\caption{Checking output for the example Dockerfile shown in \Cref{fig:scenario:file}.}\label{fig:result:example}
\end{figure}

\subsection{Context Mocking}

Due to the space limitation, we describe our design briefly.

Variables are represented by a mapping $M_{V}$ from a name to the corresponding value.
A set $S_{V}$ records variables that $M_{V}$ tracks precisely.
An extra label is used to record whether $M_{V}$ tracks every variable, denoted as $f_{V}$.

A directory is organized as a tree.
A directory is labeled as fuzzy if it fails to track every subfile.
In case that the fuzzy state of Docker container files cannot be managed by any fuzzy directory, a label is used to record it, denoted as $f_{C}$.

A list $L_{E}$ contains accessible executable names that can be searched by \texttt{PATH}.
Besides, $L_{E}$ shall include built-in utilities that are implemented by a shell interpreter, as well.
An extra label is used to record whether $L_{E}$ is fuzzy, denoted as $f_{E}$.

\subsection{Simplification}

The mock context is simplified compared with the real one.
For example, among file properties, only existence, type, and mode are taken into consideration.
File ownership is neglected.

Mock interpreters for Dockerfiles and shell scripts are also partially mocked.
While parsing a shell script, the token \textit{command} can derive \textit{simple command}, \textit{compound command}, and \textit{function definition}~\cite{POSIX:shell-grammar}.
Only the production $command : simple\_command$ is taken into consideration, as well as \textit{subshell} derived from \textit{compound command}.
\textit{function\_definition} and other tokens that are not implemented will fuzz all of the environment variables, file tree, and executable list.

\subsection{Prior Context}

In the experiment, prior context, including environment variables and the executable list, are extracted from Docker images to initialize the mock context.
Executable list contains files that can be accessed through \texttt{PATH}, as well as POSIX builtin utilities.
With prior context, $M_{V}$ can be initialized precisely, with $f_{V}$ of False.
For example, while analyzing the FROM instruction in \Cref{fig:scenario:directory}, detailed information about the Docker image can fill $L_{E}$ with basic instructions like \texttt{cd}, \texttt{npm}, and \texttt{yarn}, which are critical to a \texttt{node} image.

\section{Experiment}\label{sec:experiment}

To evaluate the performance of DockerMock, we conduct experiments on two sets of git repositories with Dockerfiles, to answer the following two research questions.

\begin{description}
    \item[\textbf{RQ2}] How effective is DockerMock on Dockerfile fault detection?
    \item[\textbf{RQ3}] How much do our assumption and prior context contribute to DockerMock's performance?
\end{description}

\subsection{Baseline Approach}

Dockerfile Linter (hadolint)~\cite{dockerfile:linter} takes a Dockerfile as input only, checks the Dockerfile and embedded shell scripts based on best practices.
We choose hadolint to represent other Dockerfile analysis tools that are also based on best practices, such as Docker extension for VS Code~\cite{vscode}.
BuildKit~\cite{buildkit} is a new backend of \texttt{docker build}.
With concurrent instruction preparation, a fault can be exposed before the instruction is executed.
Omitting actual instruction execution, BuildKit is taken as a pre-build method in the experiment.
RUDSEA~\cite{Hassan:2018} is a language-specific approach, and it is not compared in our experiment due to the implementing workload to cover all the languages in our datasets.

\subsection{Method}

\subsubsection{Labeling}\label{sec:evaluation:label}

We first apply pre-build analysis for each Dockerfile in our datasets, as shown in \Cref{fig:evaluation:faults}.
Then, we compare output warnings of a given method with labeled faults to collect the number of true positives(TP), false negatives(FN), false positives(FP).
A tool may warn the running status of itself, which should not be counted as false positives.
Meanwhile,  a reasonable suggestion will also be counted as a false positive if it is unrelated to any faults.

\begin{figure}[htb]
	\centering
	\resizebox{0.9\linewidth}{!}{
		\begin{tikzpicture}[
    startstop/.style = {rectangle, rounded corners, minimum width=1cm, minimum height=1cm,text centered, draw=black},
    process/.style={rectangle, align=center, text width=1.6cm, minimum height=1cm, text centered, draw=black},
    decision/.style={diamond, minimum width=1cm, minimum height=1cm, text centered, draw=black},
    arrow/.style={thick,->,>=stealth},
    node distance=2.5cm
    ]
    \node (start) [startstop] {Start};
    \node (init) [process, text width=3.7cm, right of=start, xshift=0.9cm] {Clone the repository};
    \node (check) [process, below of=init, yshift=0.9cm,  xshift=1.5cm] {Pre-build\\check};
    \node (build) [process, below of=init, yshift=0.9cm, xshift=-1.5cm] {\texttt{docker\\build}};
    \node (fail) [decision, below of=build, yshift=0.6cm] {Fail?};
    \node (success) [startstop, left of=fail, xshift=0.6cm] {End};
    \node (warning) [startstop, right of=check] {Warnings};
    \node (failure) [startstop, below of=check, yshift=0.6cm] {Failures};
    \node (fault) [startstop, right of=failure] {Faults};
    \coordinate (compare) at ($ (warning)!0.5! 0:(fault) $);
    \node (result) [startstop, right of=compare] {TP, FN, FP};

	\draw [arrow] (start) -- (init);
	\draw [arrow] (init.south) --+(0, -0.3cm) -| (check.north);
	\draw [arrow] (init.south) --+(0, -0.3cm) -| (build.north);
    \draw [arrow] (build) -- (fail);
    \draw [arrow] (fail) -- node[anchor=south] {N} (success);
    \draw [arrow] (fail) -- (failure);
    \draw [arrow] (fail) -- node[anchor=south] {Y} ($ (fail)!0.5! 0:(failure) $) -- node[anchor=west] {fix} ($ (build)!0.5! 0:(check) $) -- (build);
    \draw [arrow] (check) -- (warning);
    \draw [arrow] (failure) -- node[anchor=south] {classify} (fault);
    \draw [arrow] (warning) -- (compare) --  node[anchor=south] {compare} (result);
    \draw [arrow] (fault) |- (result);
\end{tikzpicture}
	}
	\caption{The evaluation workflow for labeling true positives(TP), false positives(FP), and false negatives(FN) for a given pre-build method.}\label{fig:evaluation:faults}
\end{figure}
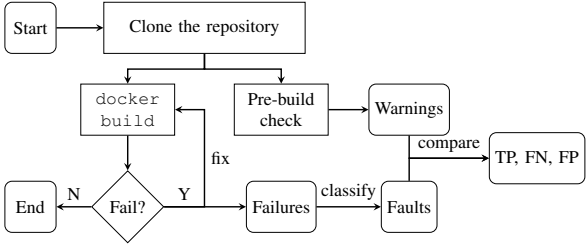


\begin{figure*}[htb]
	\centering
	\subfloat[Prospective evaluation with cache]{\label{fig:evaluation:pro:cache}
		\resizebox{0.28\linewidth}{!}{
			\begin{tikzpicture}[
    startstop/.style = {rectangle, rounded corners, minimum width=1cm, minimum height=1cm,text centered, draw=black},
    process/.style={rectangle, align=center, text width=1.6cm, minimum height=1cm, text centered, draw=black},
    decision/.style={diamond, minimum width=1cm, minimum height=1cm, text centered, draw=black},
    arrow/.style={thick,->,>=stealth},
    node distance=2cm
    ]
    \node (start) [startstop] {Start};
    \node (init) [process, text width=3.7cm, right of=start, xshift=1cm] {Clone the repository \\ Pull base Docker images};
    \node (check) [process, thick, dotted, draw=blue, below of=start] {Pre-build\\check};
    \node (checkFail) [decision, right of=check] {Fail?};
    \node (build) [process, thick, dotted, draw=blue, right of=checkFail] {BuildKit};
    \node (cache) [startstop, align=center, text width=1.1cm, right of=build] {Data\\(cache)};
    \node (buildFail) [decision, below of=build, yshift=0.3cm] {Fail?};
    \node (success) [startstop, right of=buildFail] {End};
    \node (fix) [process, below of=checkFail, yshift=0.3cm] {Fix by the researcher};

    \draw [arrow] (start) -- (init);
    \draw [arrow] (init.south) to [in = 45, out = -135] (check.north);
    \draw [arrow] (check) -- (checkFail);
    \draw [arrow] (checkFail) -- node[anchor=south] {N} (build);
    \draw [arrow] (checkFail) -- node[anchor=east] {Y} (fix);
    \draw [arrow] (build) -- (buildFail);
    \draw [arrow] (buildFail) -- node[anchor=south] {N} (success);
    \draw [arrow] (buildFail) -- node[anchor=south] {Y} (fix);
    \draw [arrow] (fix) -| (check);
    \draw[arrow, <->] (build) -- (cache);
\end{tikzpicture}
		}
	}
	\subfloat[Prospective evaluation without cache]{\label{fig:evaluation:pro:nocache}
		\resizebox{0.36\linewidth}{!}{
			\begin{tikzpicture}[
	startstop/.style = {rectangle, rounded corners, minimum width=1cm, minimum height=1cm,text centered, draw=black},
	process/.style={rectangle, align=center, text width=1.6cm, minimum height=1cm, text centered, draw=black},
	decision/.style={diamond, minimum width=1cm, minimum height=1cm, text centered, draw=black},
	arrow/.style={thick,->,>=stealth},
	node distance=2cm
	]
	\node (start) [startstop] {Start};
	\node (init) [process, text width=3.7cm, right of=start, xshift=1cm] {Clone the repository};
	\node (check) [process, thick, dotted, draw=blue, below of=start, yshift=0.2cm] {Pre-build\\check};
	\node (checkFail) [decision, right of=check] {Fail?};
	\node (build) [process, thick, dotted, draw=blue, right of=checkFail, xshift=2cm] {BuildKit};
	\coordinate (remove) at ($ (checkFail)!0.4! 0:(build) $);
	\node (cache) [startstop, align=center, text width=1.1cm, right of=build] {Data\\(cache)};
	\node (buildFail) [decision, below of=build, yshift=0.3cm] {Fail?};
	\node (success) [startstop, right of=buildFail] {End};
	\node (fix) [process, below of=checkFail, yshift=0.3cm] {Fix by the researcher};
	
	\draw [arrow] (start) -- (init);
	\draw [arrow] (init.south) to [in = 45, out = -135] (check.north);
	\draw [arrow] (check) -- (checkFail);
	\draw [arrow] (checkFail) -- (remove) --+(0, 0.8cm) node[anchor=west,yshift=0.25cm] {1. remove cache} -| (cache);
	\draw [arrow] (checkFail) -- node[anchor=south] {N} (remove) -- node[anchor=south] {2. build} (build);
	\draw [arrow] (checkFail) -- node[anchor=east] {Y} (fix);
	\draw [arrow] (build) -- (buildFail);
	\draw [arrow] (buildFail) -- node[anchor=south] {N} (success);
	\draw [arrow] (buildFail) -- node[anchor=south] {Y} (fix);
	\draw [arrow] (fix) -| (check);
	\draw[arrow, <->] (build) -- (cache);
\end{tikzpicture}
		}
	}
	\subfloat[Retrospective evaluation]{\label{fig:evaluation:retro}
		\resizebox{0.28\linewidth}{!}{
			\begin{tikzpicture}[
    startstop/.style = {rectangle, rounded corners, minimum width=1cm, minimum height=1cm,text centered, draw=black},
    process/.style={rectangle, align=center, text width=1.6cm, minimum height=1cm, text centered, draw=black},
    decision/.style={diamond, minimum width=1cm, minimum height=1cm, text centered, draw=black},
    arrow/.style={thick,->,>=stealth},
    node distance=2cm
    ]
    \node (start) [startstop] {Start};
    \node (push) [process, right of=start] {Push};
    \node (build) [process, thick, dotted, draw=blue, right of=push, xshift=0.4cm] {\texttt{docker\\build}};
    \node (fail) [decision, below of=build, yshift=0.3cm] {Fail?};
    \node (success) [startstop, right of=fail] {Delivery};
    \node (fix) [process, below of=push, yshift=0.3cm] {Fix by the developer};
    \coordinate (probe) at ($ (push)!0.5! 0:(build) $);
    \node (check) [process, above of=probe, yshift=-0.5cm] {Pre-build\\check};

	\draw [arrow] (start) -- (push);
    \draw [arrow, dashed] (probe) -- (check);
    \draw [arrow] (push) -- (build);
    \draw [arrow] (build) -- (fail);
    \draw [arrow] (fail) -- node[anchor=south] {N} (success);
    \draw [arrow] (fail) -- node[anchor=south] {Y} (fix);
    \draw [arrow] (fix) -- (push);
\end{tikzpicture}
		}
	}
	\caption{Evaluation workflows for time saved. We record how long the processes take, which are in the blue dotted frame.}
\end{figure*}
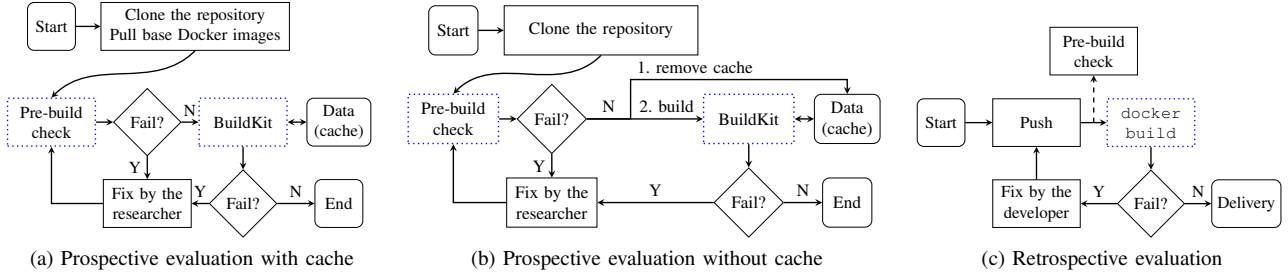

\subsubsection{Prospective evaluation}

On the GitHub dataset, the authors imitate the developer to fix the Dockerfiles in the trial-and-error approach with the help of a pre-build method, named \textit{prospective evaluation}.
Each pre-build method is applied just before invoking \texttt{docker build}.
As \Cref{sec:evaluation:label} records the fixing patches, each fault can be fixed immediately in the same way.
BuildKit does not have a standalone pre-build detection component.
We take BuildKit as the backend of \texttt{docker build}, and the ``pre-build check'' is empty for the control group with BuildKit only.
Other methods are applied with BuildKit together.
Suggestions provided by each method are taken seriously, except some given by hadolint.
For example, hadolint suggests pinning versions while we have no idea which version the developers use.
The process for each Dockerfile is executed 3 times to calculate the average for the overall execution time of checking and building.

The prospective evaluation is further adopted in two ways, compatible with the development and CI/CD environments.
A Docker image is composed of a stack of layers created by each instruction.
Layers can be reused to avoid duplicated builds, performing as \textit{cache}.
The prospective evaluation distinguishes whether the cache is utilized.
With cache, the base images are fetched while a repository is initialized, as shown in \Cref{fig:evaluation:pro:cache}.
During the build process, cache remains and accelerates the following builds.
The setting provides an approximation for the minimum time saved for a pre-build method.
Without cache, Docker data is removed before each \texttt{docker build} is invoked, as shown in \Cref{fig:evaluation:pro:nocache}.
This setting is closed to the CI/CD workflow.
And it is essential that how many times \texttt{docker build} is invoked, denoted as the \textit{number of builds}, representing the communicating and waiting cost.

\subsubsection{Retrospective evaluation}

On the student dataset, we apply what-if analysis to find out the CI failures that can be avoided by a pre-build method, named \textit{retrospective evaluation}, as shown in \Cref{fig:evaluation:retro}.
Several commits of each repository contribute to this dataset.
As the analysis cannot change the development history and the same Dockerfile fault can exist in different commits, only the first fault in a Dockerfile that triggers the CI failure is taken into consideration.
If such a fault is detected by a method $M$, it is assumed that the build time of the corresponding failure could be saved by $M$.

\subsection{Evaluation Metrics}

We first measure the effectiveness of a method $M$ by precision and recall.
While precision is computed by $\frac{TP}{TP + FP}$, recall is computed by $\frac{TP}{TP + FN}$.
The larger, the better.

Warnings shall be explainable.
hadolint has good interpret-ability as it is based on best practices, and documentation is available for its rules.
BuildKit, as well as \texttt{docker build}, provides ground truth of faults.
As a mock of \texttt{docker build}, DockerMock is as explainable as \texttt{docker build}.
The interpret-ability of DockerMock can be further measured by its precision.
Whether it is easy for a developer to understand the warnings is out of the topic in this work.

A method shall indeed save developers' time.
Supposing that fixing time is the same among different methods, a better method contributes to less time for checking and building.
In the prospective evaluation, given a method $M$, the overall execution time of \texttt{docker build} and $M$ is denoted as $T_{P}(M)$.
Time saved by $M$ is estimated as $TS_{P}(M)$, shown in \Cref{eqn:time-saved:prospective}.
Time saved rate of $M$ is estimated as $\frac{TS_{P}(M)}{T_{P}(BuildKit)}$.
\begin{equation}\label{eqn:time-saved:prospective}
	TS_{P}(M) = T_{P}(BuildKit) - T_{P}(M)
\end{equation}

In the retrospective evaluation, let $T_{R}$ be the build time of all CI failures in the student dataset.
The overall build time of failures, that can be detected by a method $M$, is assumed as the time $TS_{R}(M)$ that could be saved by $M$.
Time saved rate of $M$ is estimated as $\frac{TS_{R}(M)}{T_{R}}$.

\subsection{Results}

\subsubsection{Overall Evaluation}

The effectiveness of DockerMock is summarized in \Cref{tab:experiment:result} to answer RQ2.
In this table, we present recall and precision in each dataset. 
From this table, the recall of DockerMock is \DockerMockStudentRecall{}\% $\sim$ \DockerMockGitHubRecall{}\%, better than the other two approaches. 
The precision of DockerMock is \DockerMockGitHubPrecision{}\% $\sim$ \DockerMockStudentPrecision{}\%, quite competitive with BuildKit, and significantly better than hadolint.
In detail, we present the number of faults in each type in \Cref{tab:experiment:result:github} and \Cref{tab:experiment:result:fault:student}.
As for recall comparison, DockerMock can detect more types of faults. 
BuildKit can find the fault type of Outer-File-Not-Found, which takes up a large fraction (\StudentOuterFileNotFoundRate{} $\sim$ \GitHubOuterFileNotFoundRate{}) of all faults in two datasets.
As for precision comparison, DockerMock raises \DockerMockGitHubFalsePositive{} false positives in $\mathcal{D}_{G}$.
DockerMock warns version inconsistency between those specified by the FROM instruction and ``go.mod'' in two Dockerfiles.
The other false positive is due to \texttt{alias}, which cannot be searched by \texttt{PATH}.
In $\mathcal{D}_{S}$, there is only one false positive for \texttt{chmod}.
A warning is raised by DockerMock, as \texttt{chmod}'s \texttt{-f} option is not specified in POSIX. 
The precision of hadolint is low because it can detect most ``code smells'' rather than faults.

We further explore faults that DockerMock fails to detect.
In $\mathcal{D}_{G}$, 4 undetected faults are related to compiling or package management.
As a fix requires to update the Dockerfile, those 4 faults are also classified as Dockerfile faults.
It is worth mentioning that hadolint detects such a fault.
The failure of accessing an out-dated link is not passed through pipe, and the fault does not interrupt the build process.
7 faults are not detected due to insufficient mocks.
For example, a previous \texttt{apt} will blind DockerMock to following Command-Not-Found faults, as the implemented mock for \texttt{apt} only checks the ``-y'' option.
Since DockerMock does not utilize the network, 2 Image-Not-Found faults are not detected.
Another 3 faults are strange as \texttt{wget} is invoked, but the path of the target URL is just three dots ``...''.
In $\mathcal{D}_{S}$, except 9 Image-Not-Found faults, 35 faults are not detected due to insufficient mocks in terms of number and details.

For the time saved comparison, we compare the time saved in both $\mathcal{D}_{G}$ and $\mathcal{D}_{S}$.
In $\mathcal{D}_{G}$, pre-build checking aided trial-and-error is applied in \GitHubDatasetTimedRepoSize{} Dockerfiles, which we can fix thoroughly.
Each Dockerfile is built 3 times with cache and 3 times without cache.
In the experiment, DockerMock is executed \DockerMockGitHubExecution{} times with an average execution time of \DockerMockGitHubAvgExecutionTime{}, while hadolint is executed \hadolintGitHubExecution{} times with an average execution time of \hadolintGitHubAvgExecutionTime{}.
The execution time of DockerMock is shorter than that of hadolint statistically significantly ($p \ll 0.001$ in Wilcoxon Rank-Sum test).
With cache, DockerMock can save \DockerMockGitHubSavedTime{} out of \BuildKitGitHubBuildTime{} with a time saved rate of \DockerMockGitHubSavedTimeRate{}\%, while hadolint saves \hadolintGitHubSavedTimeRate{}\% time.
Without cache, DockerMock can save \DockerMockGitHubSavedTimeNocache{} out of \BuildKitGitHubBuildTimeNocache{} with a time saved rate of \DockerMockGitHubSavedTimeRateNocache{}\%, while hadolint saves \hadolintGitHubSavedTimeRateNocache{}\% time.
hadolint saves time mainly via shortening the duration of a single build.
For example, hadolint recommends the ``-{-}no-install-recommends'' option for \texttt{apt-get}, which lessons packages to be installed.
While DockerMock avoids \DockerMockGitHubSavedBuildTimes{}\% failed builds that are detected in a pre-build manner.

In $\mathcal{D}_{S}$, we collect the build time from \StudentDatasetTimedSize{} out of \StudentDatasetDockerfileSize{} Dockerfiles that were built in CI.
DockerMock can save detection time of \DockerMockStudentSavedTime{} out of \StudentDatasetBuildTime{} with a time saved rate of \DockerMockStudentSavedTimeRate{}\%, while hadolint and BuildKit save \hadolintStudentSavedTimeRate{}\% and \BuildKitStudentSavedTimeRate{}\% detection time, respectively.

\begin{table}[tb]
	\caption{
		Effectiveness of hadolint, BuildKit, and DockerMock.
		Build data in $\mathcal{D}_{G}$ is collected from \GitHubDatasetTimedRepoSize{} Dockerfiles, compared with BuildKit, whose build time is \BuildKitGitHubBuildTime{} with cache and \BuildKitGitHubBuildTimeNocache{} without cache.
		Sum of collected build time in $\mathcal{D}_{S}$ is \StudentDatasetBuildTime{}.
	}\label{tab:experiment:result}
	\scriptsize
	\begin{center}
		\begin{tabular}{c|l|rrr}
			\toprule
			Dataset & \multirow{2}{*}{Metrics} & \multirow{2}{*}{hadolint} & \multirow{2}{*}{BuildKit} & \multirow{2}{*}{DockerMock} \\
			(\# Faults) & & & & \\
			\midrule
			\multirow{5}{*}{$\mathcal{D}_{G}$ (\GitHubDatasetFaultSize{})} & Recall (\%) & \hadolintGitHubRecall{} & \BuildKitGitHubRecall{} & \DockerMockGitHubRecall{} \\
			& Precision (\%) & \hadolintGitHubPrecision{} & \BuildKitGitHubPrecision{} & \DockerMockGitHubPrecision{} \\
			& TSR (\%) & \hadolintGitHubSavedTimeRate{} & / & \DockerMockGitHubSavedTimeRate{} \\
			& TSR Nocache (\%) & \hadolintGitHubSavedTimeRateNocache{} & / & \DockerMockGitHubSavedTimeRateNocache{} \\
			& Number of Builds & \hadolintGitHubBuildTimes{} & \BuildKitGitHubBuildTimes{} & \DockerMockGitHubBuildTimes{} \\ \midrule
			\multirow{3}{*}{$\mathcal{D}_{S}$ (\StudentDatasetFaultSize{})} & Recall (\%) & \hadolintStudentRecall{} & \BuildKitStudentRecall{} & \DockerMockStudentRecall{} \\
			& Precision (\%) & \hadolintStudentPrecision{} & \BuildKitStudentPrecision{} & \DockerMockStudentPrecision{} \\
			& TSR (\%) & \hadolintStudentSavedTimeRate{} & \BuildKitStudentSavedTimeRate{} & \DockerMockStudentSavedTimeRate{} \\
			\bottomrule
		\end{tabular}
	\end{center}
\end{table}

\subsubsection{Contributions of Components}

To answer RQ3, we remove Assumption and prior context from DockerMock to reveal their contribution.
Faults detected by different methods in $\mathcal{D}_{G}$ and $\mathcal{D}_{S}$ are compared in \Cref{tab:experiment:result:github} and \Cref{tab:experiment:result:fault:student}, respectively.
In $\mathcal{D}_{G}$, Assumption and prior context together help DockerMock detect one more fault.
In $\mathcal{D}_{S}$, Assumption helps DockerMock detect 16 more faults without false positives, while prior context helps to find 5 more faults.

Whether a fault can be detected by DockerMock depends on at least two factors:
fault-related instruction or command is mocked;
fault-related context is certain.
A Dockerfile is sufficient to detect Syntax-Mistake and most of Instruction-Misuse.
Because the workspace is unchanged while a Docker image is under build, Outer-File-Not-Found faults always have a certain context.
Inner-File-Not-Found is another situation.
As analysis progresses, more and more context is labeled as fuzzy, which makes it harder for DockerMock to find out a fault.
The strategy makes it worse that fuzzing the whole context after a fault is detected or a mock is unavailable.
However, Assumption helps DockerMock find more such context-sensitive faults without introducing false positives in the experiment.
hadolint detects one Inner-File-Not-Found fault in $\mathcal{D}_{G}$ and two in $\mathcal{D}_{S}$ based on the best practice, using WORKDIR instead of \texttt{cd}, while DockerMock provides a systematically way to detect context-sensitive faults in a Dockerfile.
Network access is necessary to detect Image-Not-Found, which is beyond DockerMock's context.
Prior context of environment variables and executable list shows its potential to detect Command-Not-Found faults.

\begin{table}[tb]
    \centering
    \caption{\# faults detected by hadolint (HDL), BuildKit (BK), DockerMock without Assumption (DM-A), DockerMock without prior context (DM-P), and DockerMock (DM) for each fault type in $\mathcal{D}_{G}$ }\label{tab:experiment:result:github}
    \scriptsize
	\begin{minipage}[b]{\columnwidth}
\begin{tabular}{lrrrrrr}
    \toprule
    Type & Total & HDL & BK & DM-A & DM-P & DM \\
    \midrule
    Command-Misuse & 7 & 1 & 0 & 1 & 1 & 2 \\
    Command-Not-Found & 4 & 0 & 0 & 0 & 0 & 0 \\
    Outer-File-Not-Found & 32 & 0 & 32 & 32 & 32 & 32 \\
    Inner-File-Not-Found & 3 & 1 & 0 & 0 & 0 & 0 \\
    Image-Not-Found & 2 & 0 & 2 & 0 & 0 & 0 \\
    Image-Version-Mismatch & 3 & 0 & 0 & 3 & 3 & 3 \\
    Permission-Denied & 1 & 0 & 0 & 0 & 0 & 0 \\
    Require-Manual-Input & 1 & 0 & 0 & 0 & 0 & 0 \\
    \midrule
    All & 53 & 2 & 34 & 36 & 36 & 37 \\
    \bottomrule
\end{tabular}

    \end{minipage}
\end{table}

\begin{table}[tb]
	\centering
	\scriptsize
	\caption{\# faults detected for each fault type in $\mathcal{D}_{S}$} \label{tab:experiment:result:fault:student}
	\begin{minipage}[b]{\columnwidth}
\begin{tabular}{lrrrrrr}
    \toprule
    Type & Total & HDL & BK & DM-A & DM-P & DM \\
    \midrule
    Syntax-Mistake & 3 & 3 & 3 & 3 & 3 & 3 \\
    Instruction-Misuse & 8 & 4 & 3 & 7 & 7 & 7 \\
    Command-Misuse & 5 & 0 & 0 & 0 & 2 & 2 \\
    Command-Not-Found & 9 & 0 & 0 & 1 & 0 & 5 \\
    Outer-File-Not-Found & 59 & 3 & 59 & 59 & 59 & 59 \\
    Inner-File-Not-Found & 37 & 2 & 0 & 0 & 10 & 10 \\
    Image-Not-Found & 9 & 0 & 9 & 0 & 0 & 0 \\
    \midrule
    All & 130 & 12 & 74 & 70 & 81 & 86 \\
    \bottomrule
\end{tabular}

    \end{minipage}
\end{table}

\subsection{Discussion}\label{sec:discussion}

\subsubsection{False Positives}

With the fuzzy handling process, false positives may still arise.
Some main factors are listed.

\paragraph{Asynchronous execution}

As a full-featured language, the shell supports executing commands asynchronously.
%
The behavior of asynchronous execution is unpredictable and infeasible to mock.
If there is an asynchronous instruction in a Dockerfile, a false positive may occur.



\paragraph{POSIX oracle}

Shell instructions are parsed and mocked, according to POSIX.
However, shells' implementations are diverse.
The difference between the used shell interpreter/executables and our mocks can lead to false positive.




	

\paragraph{Different representation of the same path}

If two paths share the same fuzzy variable as the prefix, the relative relationship of the two paths is reserved.
Thus, our Assumption makes sense.
However, if a file is created with a fuzzy path, absolute path accessing will lead to a different file that is untouched in the previous operation.
The inverse situation may also lead to a false positive.

\subsubsection{Future Work}

The number of shell commands is a large amount and continuously increasing because shell commands can be defined by the Linux system and user software.
Therefore, it is infeasible to mock all of the shell commands manually.
As we open-source our tool, the community can also contribute to it.
Generating mocks based on documentation can help a lot.

It is necessary to discover the status of Dockerfile faults in the open-source society and private projects to verify the transferability of DockerMock.
This paper focuses on locating faults in Dockerfiles, while the proposed approach is expected to be applicable for other configuration files, \textit{e.g.}, \texttt{.gitlab-ci.yml}, which configures CI for GitLab.
Some other aspects are also meaningful. 
(1) Detecting the intention of a given Dockerfile.
(2) Repair the located Dockerfile fault, based on the detected intention.
(3) Generate Dockerfile according to the developer's intention.

\subsection{Threats to Validity}

\subsubsection{Construct Validity}

$\mathcal{D}_{S}$ is derived from student projects.
In the course, examples are prepared to show the usage of CI configuration and Dockerfile.
As the provided examples work well, Dockerfile faults are not introduced by examples.

The developing paradigm of GitHub projects is not as clear as that of student projects.
Only the Dockerfiles under the root directory of the default branch are collected.
Faults that were fixed in history are not included in our GitHub sample $\mathcal{D}_{G}$.

The reason for the same build failure may be unstable.
We built each project several times to expose the intrinsic Dockerfile faults.
Failures are fixed based on our guess of the developers' intention.
Thus, Dockerfile faults collected by different researchers may be different.
As failures and faults are classified manually, they may not be labeled correctly.

\subsubsection{Internal Validity}


It is hard to measure the time saved precisely due to the following factors: (a)A developer may fix several faults at the same time or/and introduce new ones;  (b)We failed to fix some source code/dependency faults; (c)Summing up time may enlarge errors in measurement.
In $\mathcal{D}_{S}$, evaluation is based on records collected from our CI system, which builds Docker images with cache enabled.
Those records reflect the actual fixing process by the students and provide the same set of time to compare different methods.
In $\mathcal{D}_{G}$, the authors pretend to be the developer, fixing Dockerfiles.
Though each adopted evaluation has a bias, the results of time saved with different evaluations in two datasets are consistent.

\subsubsection{External Validity}

The status of student projects is different from that in GitHub, as shown in \Cref{fig:compare:instruction}.
For some types of applications, DockerMock may get similar results.

\section{Related Work}\label{sec:related}

\paragraph{Pre-build analysis}

Dockerfile Linter (hadolint)~\cite{dockerfile:linter} checks Dockerfile and scans the script inside RUN instructions, utilizing ShellCheck~\cite{shellcheck}.
Rules of ShellCheck and hadolint are derived from best practices.
Tree association mining can be applied to identify rules from ``Dockerfiles written by Docker experts'' automatically~\cite{Henkel:2020}.
Abash focuses on the security vulnerabilities introduced by shell word expansion~\cite{Mazurak:2007}.
DockerMock is different from these tools in three aspects.
First, these tools focus on code smells, which may not block execution. While DockerMock aims at fault detection.
Second, besides the plain text of a Dockerfile, DockerMock also utilizes file structure and content of the corresponding project.
Third, these tools check based on rules and can be enhanced with constant substitution.
While DockerMock tries to reserve the semantic effect of each instruction, as well as embedded shell commands, via maintaining context.

RUDSEA utilizes the change of environment-related code scope to recommend Dockerfile update locations~\cite{Hassan:2018}.
While our approach aims to detect Dockerfile faults in-depth without touching source code.   
We will study source code and configuration files for package managers in future work.
Based on modification data and build history, machine learning can be applied to predict CI outcomes for Java projects~\cite{Hassan:2017,Chen:2020}, which is less explainable than DockerMock.

\paragraph{Context-based approach}

Free and Open Source Software distributions set up models for maintainer scripts, and maintainers specify the requirement and restriction for their scripts~\cite{DiRuscio:2009,Cosmo:2011}.
Reference \cite{Shambaugh:2016} models the instructions in an exec-free Puppet manifest and construct a resource graph to validate its determinacy.
As the embedded shell scripts do not have clear semantics, the exec resource type of Puppet is not taken into consideration in their work.
The models in \cite{DiRuscio:2009,Cosmo:2011,Shambaugh:2016} are not for Dockerfiles, while we model Dockerfile instructions and shell commands as mocks.
As Dockerfile instructions are built-in sequence, a resource graph is not necessary.
Instead, we introduce context, including environment variables and files, to bridge instructions, similar to the state table in \cite{Xu:2019}.
To handle incomplete knowledge of context, we introduce fuzzy status into context, which is different from both \cite{Shambaugh:2016} and \cite{Xu:2019}.

\section{Conclusion}\label{sec:conclusion}

Docker has been widely used in the CI/CD environment.
Dockerfile fault detection is increasingly important for developers before the build process to ensure reliability.
In this paper, we propose the first systematic study on the research problem of Dockerfile fault detection.
We present a taxonomy of Docker build faults and conduct the first empirical study on Dockerfile faults to show the need to use context for Dockerfile faults detection.
Using the context of a Dockerfile, DockerMock mocks the execution of Docker instructions and shell commands to detect Dockerfile faults in the development environment.
DockerMock achieves good accuracy in Dockerfile faults detection.
Meanwhile, DockerMock reduces the number of builds to a large extent compared to the common trial-and-error practice.
We open-source our datasets~\cite{dataset} and the code of DockerMock~\cite{code}, which can serve as benchmarks for future Dockerfile research.

\clearpage
\balance
\bibliographystyle{IEEEtran}
\bibliography{IEEEabrv,bibliography}

\begin{thebibliography}{10}
\providecommand{\url}[1]{#1}
\csname url@samestyle\endcsname
\providecommand{\newblock}{\relax}
\providecommand{\bibinfo}[2]{#2}
\providecommand{\BIBentrySTDinterwordspacing}{\spaceskip=0pt\relax}
\providecommand{\BIBentryALTinterwordstretchfactor}{4}
\providecommand{\BIBentryALTinterwordspacing}{\spaceskip=\fontdimen2\font plus
\BIBentryALTinterwordstretchfactor\fontdimen3\font minus
  \fontdimen4\font\relax}
\providecommand{\BIBforeignlanguage}[2]{{%
\expandafter\ifx\csname l@#1\endcsname\relax
\typeout{** WARNING: IEEEtran.bst: No hyphenation pattern has been}%
\typeout{** loaded for the language `#1'. Using the pattern for}%
\typeout{** the default language instead.}%
\else
\language=\csname l@#1\endcsname
\fi
#2}}
\providecommand{\BIBdecl}{\relax}
\BIBdecl

\bibitem{FowlerCI}
M.~Fowler, ``Continuous integration,''
  \url{https://martinfowler.com/articles/continuousIntegration.html}, may 2006.

\bibitem{FowlerCD}
------, ``Continuousdelivery,''
  \url{https://martinfowler.com/bliki/ContinuousDelivery.html}, may 2013.

\bibitem{Parnin:2017}
C.~{Parnin}, E.~{Helms}, C.~{Atlee}, H.~{Boughton}, M.~{Ghattas}, A.~{Glover},
  J.~{Holman}, J.~{Micco}, B.~{Murphy}, T.~{Savor}, M.~{Stumm}, S.~{Whitaker},
  and L.~{Williams}, ``The top 10 adages in continuous deployment,'' \emph{IEEE
  Software}, vol.~34, no.~3, pp. 86--95, 2017.

\bibitem{Flexera:2020}
\BIBentryALTinterwordspacing
Flexera, ``State of the cloud report,'' 2020. [Online]. Available:
  \url{https://info.flexera.com/SLO-CM-REPORT-State-of-the-Cloud-2020}
\BIBentrySTDinterwordspacing

\bibitem{Wu:2020}
Y.~Wu, Y.~Zhang, T.~Wang, and H.~Wang, ``An empirical study of build failures
  in the docker context,'' in \emph{Proceedings of the 17th International
  Conference on Mining Software Repositories}, ser. MSR '20, 2020, p. 76–80.

\bibitem{Zhang:2020}
R.~Zhang, W.~Xiao, H.~Zhang, Y.~Liu, H.~Lin, and M.~Yang, ``An empirical study
  on program failures of deep learning jobs,'' in \emph{Proceedings of the
  ACM/IEEE 42nd International Conference on Software Engineering}, ser. ICSE
  '20, 2020, p. 1159–1170.

\bibitem{dockerbuild}
\BIBentryALTinterwordspacing
{Docker Inc}, ``docker build,'' cited 2020. [Online]. Available:
  \url{https://docs.docker.com/engine/reference/commandline/build/}
\BIBentrySTDinterwordspacing

\bibitem{Cito:2017}
J.~Cito, G.~Schermann, J.~E. Wittern, P.~Leitner, S.~Zumberi, and H.~C. Gall,
  ``An empirical analysis of the docker container ecosystem on github,'' in
  \emph{Proceedings of the 14th International Conference on Mining Software
  Repositories}, ser. MSR '17.\hskip 1em plus 0.5em minus 0.4em\relax
  Piscataway, NJ, USA: IEEE Press, 2017, pp. 323--333.

\bibitem{dockerfile:linter}
\BIBentryALTinterwordspacing
L.~Martinelli, J.~L. Rodr{\'i}guez, and V.~Zeman, ``Haskell dockerfile
  linter,'' cited 2020. [Online]. Available:
  \url{https://github.com/hadolint/hadolint}
\BIBentrySTDinterwordspacing

\bibitem{Schwarz:2018}
J.~{Schwarz}, A.~{Steffens}, and H.~{Lichter}, ``Code smells in infrastructure
  as code,'' in \emph{2018 11th International Conference on the Quality of
  Information and Communications Technology (QUATIC)}, 2018, pp. 220--228.

\bibitem{Hassan:2018}
F.~Hassan, R.~Rodriguez, and X.~Wang, ``Rudsea: Recommending updates of
  dockerfiles via software environment analysis,'' in \emph{Proceedings of the
  33rd ACM/IEEE International Conference on Automated Software Engineering},
  ser. ASE 2018.\hskip 1em plus 0.5em minus 0.4em\relax New York, NY, USA: ACM,
  2018, pp. 796--801.

\bibitem{buildkit}
\BIBentryALTinterwordspacing
{Moby project}, ``Buildkit,'' cited 2020. [Online]. Available:
  \url{https://github.com/moby/buildkit}
\BIBentrySTDinterwordspacing

\bibitem{dataset}
\BIBentryALTinterwordspacing
{Anonymous}, ``Dockerfile fault datasets,'' 2020. [Online]. Available:
  \url{https://figshare.com/s/4724a117cd018096ffb0}
\BIBentrySTDinterwordspacing

\bibitem{code}
\BIBentryALTinterwordspacing
------, ``Dockermock code,'' 2021. [Online]. Available:
  \url{https://figshare.com/s/16bebc3a2d966d46ae6e}
\BIBentrySTDinterwordspacing

\bibitem{Pandit:1996}
N.~R. Pandit, ``The creation of theory: A recent application of the grounded
  theory method,'' \emph{The Qualitative Report}, vol.~2, no.~4, pp. 1 -- 15,
  1996.

\bibitem{Spadini:2017}
D.~Spadini, M.~Aniche, M.~Bruntink, and A.~Bacchelli, ``To mock or not to mock?
  an empirical study on mocking practices,'' in \emph{Proceedings of the 14th
  International Conference on Mining Software Repositories}, ser. MSR
  '17.\hskip 1em plus 0.5em minus 0.4em\relax IEEE Press, 2017, p. 402–412.

\bibitem{Regis-Gianas:2018}
Y.~R{\'e}gis-Gianas, N.~Jeannerod, and R.~Treinen, ``Morbig: A static parser
  for posix shell,'' in \emph{Proceedings of the 11th ACM SIGPLAN International
  Conference on Software Language Engineering}, ser. SLE 2018.\hskip 1em plus
  0.5em minus 0.4em\relax New York, NY, USA: ACM, 2018, pp. 29--41.

\bibitem{POSIX:shell-grammar}
{IEEE Computer Society} and {The Open Group}, Eds., \emph{IEEE Standard for
  Information Technology--Portable Operating System Interface (POSIX(R)) Base
  Specifications, Issue 7}.\hskip 1em plus 0.5em minus 0.4em\relax IEEE Std
  1003.1-2017 (Revision of IEEE Std 1003.1-2008), Jan 2018, vol.~3, ch. 2.10
  Shell Grammar, pp. 2375--2381.

\bibitem{vscode}
\BIBentryALTinterwordspacing
{Microsoft}, ``Docker for visual studio code,'' cited 2020. [Online].
  Available: \url{https://github.com/microsoft/vscode-docker}
\BIBentrySTDinterwordspacing

\bibitem{shellcheck}
\BIBentryALTinterwordspacing
V.~Holen, ``Shellcheck - a shell script static analysis tool,'' cited 2020.
  [Online]. Available: \url{https://github.com/koalaman/shellcheck}
\BIBentrySTDinterwordspacing

\bibitem{Henkel:2020}
J.~Henkel, C.~Bird, S.~K. Lahiri, and T.~Reps, ``Learning from, understanding,
  and supporting devops artifacts for docker,'' in \emph{42nd International
  Conference on Software Engineering}, ser. ICSE ’20.\hskip 1em plus 0.5em
  minus 0.4em\relax New York, NY, USA: ACM, 2020.

\bibitem{Mazurak:2007}
K.~Mazurak and S.~Zdancewic, ``Abash: Finding bugs in bash scripts,'' in
  \emph{Proceedings of the 2007 Workshop on Programming Languages and Analysis
  for Security}, ser. PLAS '07.\hskip 1em plus 0.5em minus 0.4em\relax New
  York, NY, USA: ACM, 2007, pp. 105--114.

\bibitem{Hassan:2017}
F.~{Hassan} and X.~{Wang}, ``Change-aware build prediction model for stall
  avoidance in continuous integration,'' in \emph{2017 ACM/IEEE International
  Symposium on Empirical Software Engineering and Measurement (ESEM)}, 2017,
  pp. 157--162.

\bibitem{Chen:2020}
B.~Chen, L.~Chen, C.~Zhang, and X.~Peng, ``Buildfast: History-aware build
  outcome prediction for fast feedback and reduced cost in continuous
  integration,'' in \emph{Proceedings of the 35th ACM/IEEE International
  Conference on Automated Software Engineering}, ser. ASE 2020, 2020.

\bibitem{DiRuscio:2009}
D.~Di~Ruscio, P.~Pelliccione, A.~Pierantonio, and S.~Zacchiroli, ``Towards
  maintainer script modernization in foss distributions,'' in \emph{Proceedings
  of the 1st International Workshop on Open Component Ecosystems}, ser. IWOCE
  '09.\hskip 1em plus 0.5em minus 0.4em\relax New York, NY, USA: ACM, 2009, pp.
  11--20.

\bibitem{Cosmo:2011}
R.~D. Cosmo, D.~D. Ruscio, P.~Pelliccione, A.~Pierantonio, and S.~Zacchiroli,
  ``Supporting software evolution in component-based foss systems,''
  \emph{Science of Computer Programming}, vol.~76, no.~12, pp. 1144 -- 1160,
  2011, special Issue on Software Evolution, Adaptability and Variability.

\bibitem{Shambaugh:2016}
R.~Shambaugh, A.~Weiss, and A.~Guha, ``Rehearsal: A configuration verification
  tool for puppet,'' in \emph{Proceedings of the 37th ACM SIGPLAN Conference on
  Programming Language Design and Implementation}, ser. PLDI ’16.\hskip 1em
  plus 0.5em minus 0.4em\relax New York, NY, USA: Association for Computing
  Machinery, 2016, p. 416–430.

\bibitem{Xu:2019}
J.~{Xu}, Y.~{Wu}, Z.~{Lu}, and T.~{Wang}, ``Dockerfile tf smell detection based
  on dynamic and static analysis methods,'' in \emph{2019 IEEE 43rd Annual
  Computer Software and Applications Conference (COMPSAC)}, vol.~1, Jul 2019,
  pp. 185--190.

\end{thebibliography}

\end{document}